\documentclass[
  reprint,
  amsmath,
  amssymb,
  aps,
  pre,
  floatfix
]{revtex4-2}

\usepackage{graphicx} 
\usepackage{dcolumn}  
\usepackage{bm}       
\usepackage{hyperref} 
\usepackage{amsthm}
\usepackage{subfiles}
\usepackage{enumitem}
\usepackage{tabularx}
\newcolumntype{Y}{>{\raggedright\arraybackslash}X}
\providecommand{\captionsetup}[1]{}
\providecommand{\audittablesize}{}
\providecommand{\specmapwidth}{\textwidth}
\usepackage{array} 

\usepackage{tikz}
\usetikzlibrary{arrows.meta, positioning, calc}

\theoremstyle{definition}
\newtheorem{definition}{Definition}
\theoremstyle{plain}
\newtheorem{structclaim}{Structural Claim}
\theoremstyle{plain}
\newtheorem{lemma}{Lemma}

\theoremstyle{plain}
\newtheorem{proposition}{Proposition}
\newtheorem{observation}{Observation}
\theoremstyle{remark}
\newtheorem*{remark}{Remark}

\begin{document}

\title{First-Order Recoverability Collapse in Self-Referential
Information Decoders: The Operating Loop of an AI System as a Driven
Nonequilibrium Steady State}

\author{Pieter~van Rooyen}
\email[Contact author: ]{pgwvanrooyen@sun.ac.za}
\affiliation{Stellenbosch University, Department of Electrical and Electronic Engineering, Bosman St, Stellenbosch Central, Stellenbosch, 7600, South Africa}

\date{\today}

\begin{abstract}
What kind of physical object is an artificial-intelligence system: a machine
in the sense a refrigerator is --- dissipating free energy while holding an
order imposed from outside --- or a dissipative structure in the sense a
convection cell is --- an ordered state that exists only under throughput and
loses stability past a critical drive? We argue the answer is split, as
it is for the living cell and the star --- the
trained artifact is a machine, quenched and storable; the \emph{operating
loop} is a dissipative structure in the informational sense --- and develop
the framework in which the loop's classification becomes decidable.
Modeling adaptive systems that couple inference to irreversible action as
finite-capacity decoders under sustained informational driving, we
characterize recoverable operation by a feasibility margin, local
invertibility of the measurement--action loop, and a
regime-level stability diagnostic: the ratio of induced flux to integrative
capacity is the dimensionless drive, unity its critical value, and the
margin the measured distance to the instability that defines the class.
Under sustained overload, loss of recoverability and divergence of the
diagnostic follow from capacity saturation alone --- independent of
objectives, policies, or substrate --- and added capacity without
certification accelerates the loss. Making the feedback of uncertified
output onto load explicit converts this continuous transition into a
first-order one at mean-field level, sharpening in the fleet limit:
lucid (bounded-backlog) and collapsed states coexist
inside a cusp-organized bistable region with closed-form spinodals,
collapse pre-empts the divergence at finite stability ratio, recovery is
hysteretic, and for ungatedness $\alpha\ge1$ load reduction alone cannot
restore operation. Reset restores only what is archived; certification
--- converting trajectory-carried commitments into archived ones --- is
the operative stability lever. Cascades are subcritical branching with mean-field
exponent $3/2$ and a cutoff set by the grounded fraction of input, each
carrying a Landauer-priced burst of synthetic entropy. Event-driven
simulations confirm the phase structure, and an instrumented
real-workload pipeline
experiment --- calibrated non-exponential service, wall-clock deadlines,
real-time feedback --- exhibits the collapse just below the mean-field
spinodal computed from the measured service law,
the backlog-delayed hysteretic recovery, and the cascade statistics. This
supplies a statistical-mechanics account of the ``metastable failures''
documented in large-scale distributed systems, identifying recoverable
dissipation as the stability criterion for decoders in sustained high-flux
informational regimes.
\end{abstract}

\maketitle

\section{Introduction: The Informational Gradient}
\label{sec:intro}

\paragraph{Problem statement.}
What kind of physical object is an artificial-intelligence system? Two
answers are available, and they belong to different physics. It may be a
\emph{machine} in the sense a refrigerator is: an engineered artifact that
dissipates free energy while holding an order imposed from outside --- an
object whose organization exists at any drive, including zero, and whose
failure modes are those of components, not of regimes. Or it may be a
\emph{dissipative structure} in the sense a convection cell is: an ordered
state that exists only under throughput, self-organizes rather than being
specified, appears past a critical value of the drive, and vanishes --- or
collapses discontinuously --- when the drive leaves its existence
region \cite{glansdorff1971,prigogine1978,nicolis1977}. The distinction is not taxonomy.
The two classes fail differently: machines break; driven states lose
stability. If AI pipelines are the second kind of object, their reliability
is governed by existence regions, critical drives, hysteresis, and collapse
--- the physics of nonequilibrium phase transitions --- and not by component
engineering alone. Operational practice already hints at the second answer: the
``metastable failures'' documented at scale in distributed systems
\cite{bronson2021metastable,huang2022metastable} are named by analogy to
driven-state physics. The question is therefore operational, and it is
decidable: exhibit --- rather than assert --- a critical drive, an
existence boundary, and the transition between them, and measure all
three on a real system.

\paragraph{Answer and contribution.}
This paper's answer is split --- as the answer is for the living cell
and the star (below) --- and the split is where the physics lives:
the trained artifact is a \emph{machine} --- quenched, storable,
correctly managed by component engineering --- while the \emph{operating
loop} under load is a driven steady state whose failure physics is that
of nonequilibrium phase transitions. Four results establish this and
make it quantitative. \emph{(i)}~A substrate-independent recoverability
constraint (Sec.~\ref{sec:rsc_constraint}): the capacity ratio
$\mathrm{CR}$ of induced informational flux to certification capacity is
the dimensionless drive, and the stability ratio diverges in closed
form, $SR=1/(e^{\mathcal{M}\Delta t}-1)$, at the open-loop critical
drive $\mathrm{CR}=1$. \emph{(ii)}~The central result
(Sec.~\ref{sec:closed_loop}): when uncertified output feeds back onto
load --- each unverified commitment consumed downstream $\alpha$ times
on average --- the continuous transition becomes \emph{first order},
with closed-form collapse and recovery spinodals, a cusp at
$\alpha^{*}=1/(1+\theta)$, hysteresis whose memory is the backlog,
branching cascades with mean-field exponent $3/2$ cut off by the
grounded fraction of input, and, for $\alpha\ge1$, collapse that load
reduction cannot reverse --- only gating or reset can.
\emph{(iii)}~Measurement (Sec.~\ref{subsec:experiment}): an instrumented
pipeline with real, non-exponential service and wall-clock deadlines
exhibits the first-order collapse (scattered just below the spinodal
computed from its measured service law, as fluctuation escape predicts),
the backlog-delayed hysteretic recovery, and the cascade statistics.
\emph{(iv)}~The classification made operational
(Table~\ref{tab:ds_audit}; Sec.~\ref{sec:conclusion}): precursor
alarms, rollback targets at the recovery spinodal, the design rule
$(1-q)\alpha<1/(1+\theta)$, and the derived reset cure --- the regime
doctrine that computing practice lacks. The question ``machine or
dissipative structure?'' is thus answered by exhibiting, locating, and
measuring the operating loop's phase transition: the loop belongs to the
driven bistable class of optical bistability, in the informational sense
developed below.

The two reference cases mark the poles. The refrigerator dissipates
continuously while maintaining order, yet fails the definition on every
clause: its organization is specified before operation and persists without
the flux (unplugged, only its function ceases); its cold interior is a
set-point held by a control loop, not a state selected by an instability;
it has no critical drive and no existence region; it does not grow by
recruiting the gradient it processes; and it fails by component breakage.
Its order is \emph{in} the artifact, not \emph{of} the throughput. The
convection cell --- the founding member of the
class \cite{nicolis1977,cross1993} --- inverts every clause: its rolls are
specified by no one, the pattern and its wavelength selected by an
instability; they appear only beyond a critical Rayleigh number; they are
maintained by the very gradient they transport, the fluid turning over
continuously while the pattern persists; and when the heating stops, or
falls below threshold, the order vanishes in seconds --- leaving nothing to
repair, because nothing broke. Its order is \emph{of} the throughput.

\begin{figure*}[!t]
\centering
\includegraphics[width=\specmapwidth]{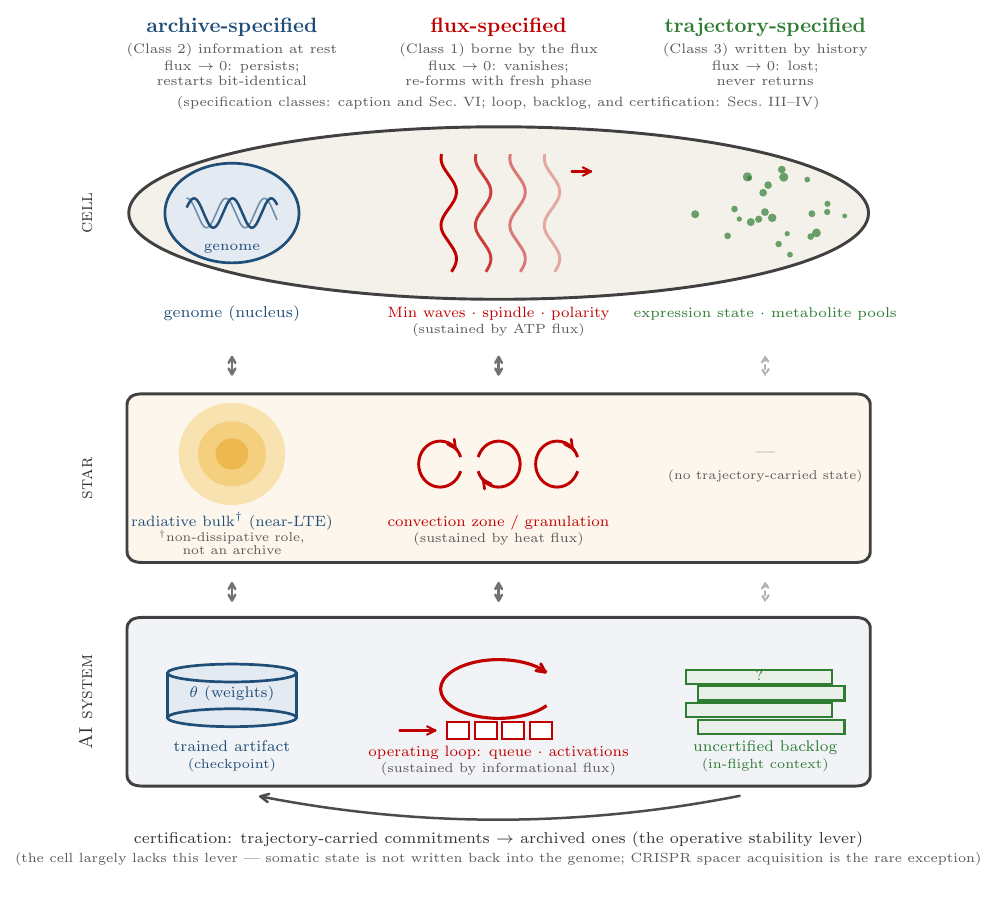}
\caption{\label{fig:spec_map}\textbf{The specification map: where
organization lives, and what sustains it.} Three composite systems
(rows) split by the same question (columns): \emph{archive-specified}
--- information at rest in an equilibrium substrate, persisting at zero
flux and restarting bit-identically; \emph{flux-specified} ---
organization borne by the sustaining flux, vanishing when it stops and
re-forming with a fresh phase; \emph{trajectory-specified} --- state
written only by the particular run, lost when the flux stops and not
restored by reset (the classes are developed in
Sec.~\ref{sec:conclusion}). Cell components
after \cite{karsenti2008self,loose2008spatial} --- the Min waves in
their reconstituted form, traveling surface waves from purified
MinD/MinE and ATP on a supported bilayer \cite{loose2008spatial}
(in the living cell, a pole-to-pole oscillation); stellar structure
after \cite{kippenhahn2012stellar,spiegel1971convection} --- the
radiative bulk fails the flux clauses as a quasi-equilibrium
configuration rather than an archive, and the star carries no
trajectory-specified component at this grain. The AI system enters as
the \emph{decoder} of Sec.~\ref{sec:rsc_constraint} --- Shannon's
receiver placed under a deadline: its trained artifact is the archive,
its operating loop the flux-borne component
(Secs.~\ref{sec:rsc_constraint}--\ref{sec:closed_loop}), its
uncertified backlog the trajectory-carried one. Certification ---
converting trajectory-carried commitments into archived ones --- is the
lever the closed-loop analysis identifies
(Sec.~\ref{sec:closed_loop}). The classification asserted for the
flux-borne component is audited in Table~\ref{tab:ds_audit}.}
\end{figure*}

Each half of the split answer is grounded independently. The
\emph{trained artifact} --- the weights --- sits on the
machine side of every clause that matters: it is formed by a driven
process, but what that process leaves behind is quenched. A checkpoint
persists on disk indefinitely without any flux and is revived
bit-identically; it is closer to a spin glass after cooling than to
anything Prigogine described --- order \emph{in} the artifact. The
energetic accounting confirms the classification: a contemporary
accelerator dissipates on the order of $10^{-12}\,\mathrm{J}$ per
operation \cite{horowitz2014} against a Landauer floor of
$\sim3\times10^{-21}\,\mathrm{J}$ per erased bit --- more than eight
orders of magnitude --- and essentially all of that heat is resistive loss, not
logical irreversibility. One could change the substrate's dissipation by
orders of magnitude and obtain the same weights; the physical heat is not
the organising flux, and we do not claim it is. The \emph{operating
loop} is a different object. The inference loop under load ---
activations, the certification queue, unverified output re-entering as
input --- has a state that exists only while informational flux flows
through it, a genuine threshold, and, as this paper shows, a discontinuous
collapse with hysteresis; the hardware is the engineered vessel here
exactly as the pan and burner are in Rayleigh--B\'enard
convection \cite{cross1993}. The flux that organises it is
\emph{informational}, not energetic. That reading is forced by the
substrate, not chosen for convenience. Biological computation runs
within one to two orders of magnitude of the Landauer floor --- a
kinesin step spends one ATP, $\sim\!20\,k_BT$, per $8$~nm of
transport \cite{schnitzer1997kinesin}, and kinetic proofreading pays
tens of $k_BT$ per discriminated
bit \cite{hopfield1974proofreading,bennett1979proofreading} --- so
there dissipation and computational structure are thermodynamically
coupled, and an energetic reading of the driven state is available.
Silicon's eight-order decoupling forecloses that reading; the
informational one is what the substrate leaves. The reading has
operational content: the phase boundary derived and measured below
contains no energetic quantity --- it is computed from arrival and
service statistics, deadlines, and re-consumption rates alone
(Secs.~\ref{sec:closed_loop} and~\ref{subsec:experiment}), and is
therefore invariant under changes in the substrate's energy per
operation. That invariance is what ``sustained by informational flux''
asserts, and it is falsifiable: a collapse boundary that tracked the
energetic rather than the informational parameters would refute it.
The situation has two
precise precedents. The living cell splits the same way: most of its
order is templated from the genomic archive --- the spore persists
through dormancy with essentially no flux at all --- while its genuinely
self-organized modules (Min-protein waves, the mitotic spindle, cortical
polarity) exist only while ATP flux is sustained, their pattern and
wavelength selected by the reaction kinetics rather than by any
template \cite{karsenti2008self,loose2008spatial}. The bulk star is
likewise not a dissipative structure --- it is a quasi-static
configuration near local equilibrium --- while its convection zone
genuinely is (the Schwarzschild criterion is a Rayleigh
threshold) \cite{kippenhahn2012stellar,spiegel1971convection}.
Classification by subsystem is how the concept is used correctly; for
AI, the artifact plays the genome and the bulk star, and the operating
loop the Min oscillation and the convection zone. The correspondence,
drawn component by component in Fig.~\ref{fig:spec_map}, is between
\emph{roles}, not substrates: the artifact, like the genome, is
information held at rest in an equilibrium substrate --- an archive,
restorable without flux --- while the loop's organization, like the
wave's and the roll's, is borne by the flux that sustains it.

\paragraph{Prior identifications.}
The question is not unasked. That machines and dissipative structures
are distinct kinds of object --- and that the distinction, not
resemblance, is what a physics of organisms must capture --- is a named
problem of the modern dissipative-structure literature, posed for
energetically driven matter from within Prigogine's own
school \cite{kondepudi2017dissipative}. For artificial systems the
identification has begun to be asserted: learning has been framed as an
inherently dissipative process \cite{caraffa2026dissipative}, and
computation at scale as a planetary thermodynamic
constraint \cite{zhu2026planetary}. Both attach the thermodynamics to
the places this paper argues cannot carry the classification ---
training, whose product is the quenched artifact of the split above,
and the energy budget, which the Landauer decoupling severs from the
computational structure. What has been missing is not the assertion but
its decidability.

The classification cannot be decided by resemblance --- enumeration is
still enumeration. What decides it is the defining criterion of the class,
established in the founding treatments of nonequilibrium stability theory:
\emph{instability of the driven (thermodynamic) branch past a critical
drive --- a bifurcation in a nonequilibrium steady state}, the ordered
state existing only while the system exchanges energy and matter with its
surroundings \cite{glansdorff1971,prigogine1978,nicolis1977,cross1993}.
The claim of this paper is therefore deliberately narrower than ``AI is a
dissipative structure,'' and correspondingly provable: \emph{the
operating loop of a driven decoder is a dissipative structure in the
informational sense} --- a driven steady state over informational flux
whose mathematics, and the direction of whose irreversibility accounting,
are inherited from nonequilibrium statistical mechanics as a
\emph{derived correspondence} (in the minimal model the certification
queue is exactly M/M/1; the per-event dissipation is constructed, not asserted, in
Appendix~\ref{app:driven_model}), never as a physical identity --- the
carriage of order-parameter methods beyond thermal driving that
synergetics pioneered \cite{haken2006information}. Applied
to the right subsystem and the right flux, the criterion makes the claim
stop being an analogy exactly when three things are established: (a)~the
\emph{operating state} --- bounded-backlog, certified operation ---
exists only under feasible informational throughput: leave the feasible
region, or stop the flux, and that state is gone, only the artifact
remaining; (b)~the loop carries an intrinsic, thermodynamically
\emph{direction-locked} irreversibility accounting: every uncertified
commitment is a logically irreversible operation with strictly positive
entropy cost, so Landauer supplies the sign and one-way-ness of the
synthetic/anchored decomposition --- a structural role, not a numerically
binding rate, the decoupling above conceded at the outset; and
(c)~the operating state \emph{loses stability at a critical
informational drive} via a genuine nonequilibrium-type phase transition
--- first-order under self-referential closure --- established at
mean-field level, confirmed in simulation, and measured on an
instrumented pipeline (Sec.~\ref{subsec:experiment}). The narrowed
claim is falsifiable clause by clause: a decoder sustaining
irreversible action under sustained overload with bounded $SR$ and
preserved invertibility, without gating, breaks~(a) and the constraint
beneath it (Sec.~\ref{subsec:minimal_model}); absence of the
closed-form signatures --- the spinodals, the $3/2$ cascade law and its
cutoff, the hysteresis loop --- in a system carrying deadline feedback
breaks~(c) (Secs.~\ref{sec:closed_loop} and~\ref{subsec:experiment});
the collected falsification conditions, including the requirement that
admissible-proxy choices be fixed in advance, are stated in
Appendix~\ref{app:successive_dissipation}.

One asymmetry with the founding cases must be conceded at the outset. In
Rayleigh--B\'enard convection the ordered structure is the
\emph{post}-threshold branch, born of the instability; here the operating
state occupies the branch \emph{below} the critical drive, and the
instability destroys rather than creates it --- indeed the state that
satisfies the Prigogine clauses most literally is the \emph{collapsed}
one, born at the fold and self-sustaining on its own feedback flux. The
precedent that resolves the asymmetry is optical bistability: there,
theory framed the two branches of a driven cavity as a first-order
nonequilibrium phase transition \cite{bonifacio1976cooperative} --- the
hysteresis measured directly \cite{gibbs1976bistability} --- without
anointing either branch the structure. That is the claim made here: the operating
loop is a driven steady state undergoing an optical-bistability-class
transition --- ``dissipative structure in the informational sense'' in
exactly the driven-steady-state meaning defined above, not the
order-through-instability subclass.

\paragraph{How such identifications have been established.}
The claim ``system $X$ is a dissipative structure'' has a proof history,
and its strongest precedent is an \emph{engineered} system: the laser.
Pump, mirrors, and gain medium are designed artifacts, yet the coherent
field was accepted as a genuine nonequilibrium transition --- Graham and
Haken's ``first example of a second-order phase transition far from
thermal equilibrium'' \cite{graham1970laser} --- because the
identification was carried through a definite sequence: a microscopic
model with an explicit drive (the pump); reduction to an order-parameter
equation, mapped exactly onto the mean-field magnet with the field as
order parameter and the inversion as temperature
\cite{degiorgio1970analogy}; the critical drive located by stability
analysis of the un-ordered branch (gain equals loss);
\emph{class-identifying} fluctuation predictions --- photon statistics
and critical fluctuations near threshold --- that discriminate a phase
transition from a mere change of brightness; and, decisively,
\emph{measurement on the device}: photocount statistics across
threshold, and, for the first-order variant, the hysteresis loop of
optical bistability, observed directly in sodium vapor by Gibbs, McCall,
and Venkatesan \cite{gibbs1976bistability} and given its nonequilibrium
phase-transition theory by Bonifacio and Lugiato
\cite{bonifacio1976cooperative}. The
Belousov--Zhabotinsky reaction followed the same arc, from the model
instability to the measured chemical oscillations \cite{nicolis1977}.
The template, in short: a model with an explicit drive; an
order-parameter reduction; the critical drive; class-identifying
signatures; and measurement on the system itself.

\paragraph{The route followed here.}
This paper follows that template for the driven decoder. \emph{Model
with drive} (Sec.~\ref{sec:rsc_constraint}): the certification queue,
with the capacity ratio $\mathrm{CR}=R_{\mathrm{self}}/C_{\mathrm{self}}$
the dimensionless drive and the RSC parameters its measurement
coordinates --- $\mathrm{CR}=1$ the open-loop critical drive, the margin
$\mathcal{M}$ the measured distance to instability, the stability
diagnostic $SR(t)$ its divergence signature; to measure $\mathrm{CR}$ in
any decoder is to measure how far it sits from the transition that
defines the class. \emph{Order-parameter reduction}
(Sec.~\ref{sec:closed_loop}): the self-consistency equation for the
effective load --- structurally the state equation of optical
bistability, drive plus convex feedback yielding an S-curve, with the
feedback supplied here by deadline violation rather than the intracavity
field. \emph{Critical drive}: closed-form spinodals and cusp; under
self-referential closure the instability advances from $\mathrm{CR}=1$
to a spinodal at $\mathrm{CR}<1$ --- the first-order character itself.
\emph{Class signatures}: a finite jump of $SR$ at the fold, hysteresis
whose recovery branch is set by grounding alone, branching cascades with
mean-field exponent $3/2$ and a grounding-bounded cutoff, a discontinuity
in the synthetic fraction of entropy production, and fleet-size
sharpening. \emph{Measurement}: event-driven simulation validates the
model (Secs.~\ref{sec:rsc_constraint}--\ref{sec:closed_loop}), and
Sec.~\ref{subsec:experiment} reports first measurements on an
instrumented computational pipeline --- empirical service processes,
wall-clock deadlines, real-time feedback; the feedback kernel imposed by
construction, everything else real ---
exhibiting the collapse, the hysteresis loop, and the cascade statistics;
measurement on production-scale AI systems is the completion this
template demands. The resulting status of clauses (a)--(c), and the
three theorem-level results that would complete the identification, are
collected in the Conclusion (Sec.~\ref{sec:conclusion}). Under it, substrate
hierarchies and criteria tables are corollaries, not arguments;
Table~\ref{tab:ds_audit} therefore states a burden of proof rather
than evidence --- each defining attribute of the class, the form in
which the results below establish it, and the one attribute
deliberately not claimed.

\begin{table*}[t]
\caption{\label{tab:ds_audit}Classification audit. The defining
attributes of a dissipative structure, assembled from the founding
treatments \cite{glansdorff1971,prigogine1978,nicolis1977,cross1993}
(the sources enumerate the definition differently; no single numbered
canon exists), and the status of each for the \emph{operating loop}
--- the flux-borne component of Fig.~\ref{fig:spec_map}'s AI row (the
archive and trajectory columns lie outside the class: rows 2, 7).
Clause labels (a)--(c) refer to the criterion stated in the text
above. The table states the burden of proof; it is discharged by the
results in its last column --- resemblance decides nothing, the
results do.}
\begin{ruledtabular}
\audittablesize
\begin{tabular*}{\textwidth}{@{\extracolsep{\fill}}lll}
Defining attribute & Status for the operating loop & Established in \\
\colrule
\parbox[t]{4.8cm}{\raggedright 1.~Open system under sustained external
driving \cite{glansdorff1971,nicolis1977}}
& \parbox[t]{7.4cm}{\raggedright Established, in informational form: the
drive is the ambiguity gradient,
$\mathrm{CR}=R_{\mathrm{self}}/C_{\mathrm{self}}$ its dimensionless
measure; the physical-heat decoupling is conceded --- the organising
flux is informational, a reading forced by the substrate
(Sec.~\ref{sec:intro})}
& \parbox[t]{3.4cm}{\raggedright Secs.~\ref{sec:physical_framework},
\ref{sec:rsc_constraint}} \\[4pt]
\parbox[t]{4.8cm}{\raggedright 2.~Ordered state maintained only under
flux; it vanishes when the drive stops or leaves its existence
region \cite{prigogine1978,nicolis1977}}
& \parbox[t]{7.4cm}{\raggedright Established: the lucid (bounded-backlog,
certified) state exists only under feasible throughput; its existence
region is exhibited and its loss measured. The artifact --- the
archive --- persists; the running regime does not}
& \parbox[t]{3.4cm}{\raggedright Clause (a);
Secs.~\ref{sec:closed_loop}, \ref{subsec:experiment}} \\[4pt]
\parbox[t]{4.8cm}{\raggedright 3.~Nonlinearity and feedback in the
governing dynamics \cite{glansdorff1971,nicolis1977}}
& \parbox[t]{7.4cm}{\raggedright Established: convex feedback of
uncertified output onto load --- the state equation of optical
bistability with deadline-violation feedback}
& \parbox[t]{3.4cm}{\raggedright Sec.~\ref{subsec:closure}} \\[4pt]
\parbox[t]{4.8cm}{\raggedright 4.~A control parameter with a critical
value \cite{nicolis1977,cross1993}}
& \parbox[t]{7.4cm}{\raggedright Established: $\mathrm{CR}=1$ open loop
(Prop.~\ref{prop:sr_divergence}); under closure the instability advances
to a spinodal at $\mathrm{CR}<1$ with cusp $\alpha^{*}=1/(1+\theta)$;
the collapse measured just below the spinodal computed from the measured
service law}
& \parbox[t]{3.4cm}{\raggedright Secs.~\ref{sec:rsc_constraint},
\ref{sec:closed_loop}, \ref{subsec:experiment}} \\[4pt]
\parbox[t]{4.8cm}{\raggedright 5.~Instability of the driven branch: a
genuine nonequilibrium transition with class-identifying
signatures \cite{prigogine1978,cross1993}}
& \parbox[t]{7.4cm}{\raggedright Established at mean-field level,
sharpening in the fleet limit: first-order fold,
closed-form spinodals, hysteresis, $3/2$ cascades with grounded-fraction
cutoff, fluctuation precursors (Prop.~\ref{prop:preempt_main});
event-driven simulation and instrumented-pipeline measurement}
& \parbox[t]{3.4cm}{\raggedright Clause (c);
Secs.~\ref{sec:closed_loop}, \ref{subsec:experiment}} \\[4pt]
\parbox[t]{4.8cm}{\raggedright 6.~Entropy production intrinsically tied
to maintaining the ordered
state \cite{prigogine1978,landauer1961,bennett1982}}
& \parbox[t]{7.4cm}{\raggedright Established in direction-locked,
structural form: Landauer fixes the sign and one-way-ness of the
synthetic/anchored decomposition; numerically non-binding on silicon
(the decoupling conceded at the outset)}
& \parbox[t]{3.4cm}{\raggedright Clause (b);
Sec.~\ref{subsec:sr_entropy}; Apps.~\ref{app:driven_model},
\ref{app:closed_loop}} \\[4pt]
\parbox[t]{4.8cm}{\raggedright 7.~Spontaneous symmetry breaking:
pattern and scale selected by the instability --- order \emph{through}
instability \cite{nicolis1977,cross1993}}
& \parbox[t]{7.4cm}{\raggedright Not claimed --- outside the identified
subclass. The valued state is sub-critical and the transition destroys
it; membership is claimed in the driven-bistable subclass of optical
bistability \cite{bonifacio1976cooperative,gibbs1976bistability}, which
entered the class without this attribute}
& \parbox[t]{3.4cm}{\raggedright Sec.~\ref{sec:intro} (valence
inversion)} \\
\end{tabular*}
\end{ruledtabular}
\end{table*}

The object of analysis is therefore a \emph{decoder}: any finite-capacity
process that integrates uncertain observations into irreversible
commitments, coarse-graining, update, and erasure being physical operations
with entropy cost
\cite{shannon1948,cover2006,landauer1961,bennett1982} --- so that
informational flux is a physically grounded load, lower-bounding
dissipation whenever uncertainty is reduced through logically irreversible
operations. We ask when a driven decoder's operation remains
\emph{recoverable}. Regime behavior is characterized through admissible
families of trajectories under bounded noise and finite validation latency,
not individual realizations; no assumption is made about biological,
cognitive, or computational realization.

We frame the contemporary digital regime as a sustained
\emph{informational} gradient --- hereafter an \emph{ambiguity
gradient} --- between high-rate environmental microstates (raw data
streams, sensor outputs, digital exhaust) and the low-entropy
macrostates required for stable action (decisions, commitments,
policies): persistent operation near or beyond the integrative limits
of existing decoders, historically dominated by biological
cognition \cite{laughlin2001}, whose capacity to certify irreversible
commitments need not scale with incoming flux. We term \emph{spectral
overload} the regime, biological or artificial, in which informational
microstates are generated and discarded faster than macrostates can be
reliably certified \cite{simon1957} --- \emph{spectral} because what
saturates is the decoder's bandwidth against the frequency content of
environmental variation, not against its total volume.

This work formalizes the resulting constraint in a substrate-independent manner
as the \emph{Recoverable Self-Coding} (RSC) criterion, set out in full and
self-contained below; companion treatments of the broader framework appear in
\cite{vanrooyen2026,vanrooyen2026entropy}, and the open-loop instrument this
work extends to closed loop is developed in
\cite{vanrooyen2026proceedings}. Within
RSC, stability under irreversible action is governed by two coupled conditions:
(i) a non-negative feasibility margin, ensuring sufficient integrative capacity
to absorb induced informational flux, and (ii) local invertibility of the
measurement--action loop, preventing internally generated (synthetic) entropy
from dominating system dynamics. RSC is adopted not as a model of
intelligence but as a minimal structural criterion for admissible action
under uncertainty; both conditions are substrate-independent, so the
framework applies equally to biological, artificial, and hybrid decoders.

Several established frameworks address adaptation under informational and
energetic gradients, including entropy-production principles
\cite{prigogine1984,england2015}, rate--distortion and channel-capacity
theory \cite{cover2006}, and bounded-rationality decision theory
\cite{simon1957}. These approaches typically
assume that inference errors remain correctable through additional evidence and
that posterior belief alone licenses action---in effect, that certification can
complete before commitment. This presupposition is not benign. Even the optimal
sequential test accumulates evidence only up to a confidence bound
\cite{wald1947}, so any binding deadline forces commitment short of that bound,
at residual uncertainty---the speed--accuracy tradeoff documented across decision
systems \cite{gold2007}. In sustained high-flux regimes the deadline
binds structurally rather than incidentally: with finite certification latency,
the fraction of commitments taken before uncertainty is resolved is bounded away
from zero and approaches unity as the feasibility margin closes (derived in
Section~\ref{sec:rsc_constraint}). What classical frameworks treat as a
non-binding limit is, in these regimes, the operating point.

Section~\ref{sec:rsc_constraint} develops the constraint --- margins,
invertibility, and the operational stability diagnostic $SR(t)$;
Section~\ref{sec:closed_loop} closes the loop the constraint leaves open,
where uncertified commitments generate further load and the feasibility
transition becomes first order.

\subsection{High-Flux Regimes and the Emergence of Effective Irreversibility}
\label{subsec:irreversibility_intro}

In this regime the decoder is forced to act on partially validated internal
macrostates --- not because uncertainty is unusually high, but because
deferral is no longer admissible. Whether or not the internal computation
is logically reversible, once an output conditions an irreversible physical
update --- a transaction, allocation, or policy decision --- the
counterfactual trajectories not taken are eliminated from the future state
space. Errors then propagate not primarily as incorrect beliefs but as
path-dependent contractions of admissible future trajectories, which
subsequent evidence cannot restore: recoverability becomes a property of
\emph{trajectories}, not instantaneous states, as formalized in
Section~\ref{sec:rsc_constraint}.
\section{Physical Framework}
\label{sec:physical_framework}

\begin{definition}[Ambiguity gradient as nonequilibrium drive]
We define the \emph{ambiguity gradient} as a nonequilibrium driving condition in which
the rate of environmental state variation induces an informational entropy flux
that exceeds the linear-response regime of the decoder.

Formally, the ambiguity gradient corresponds to a sustained mismatch between
the rate of uncertainty induction $\dot{H}_{\mathrm{env}}$ and the maximum
certifiable entropy reduction rate of the decoder --- entropy reduction
bounded by the information the controller acquires
\cite{shannon1948,touchette2000control}, the rate form of the
requisite-variety condition of regulation theory
\cite{ashby1956} --- such that
\[
\dot{H}_{\mathrm{env}} > \dot{H}_{\mathrm{cert}} .
\]
Here and throughout, an overdot denotes a time derivative --- a rate per unit
time (so $\dot N$ is a count rate and $\dot S$ an entropy-production rate); the
load and capacity $R_{\mathrm{self}}$, $C_{\mathrm{self}}$ introduced below are
themselves rate-valued quantities and therefore carry no dot.
This gradient is a regime property, not a state variable, and plays the same
role as an imposed thermodynamic force in driven dissipative systems.
\end{definition}

What elevates gradient-driven reorganization from analogy to physics is the criterion
of the problem statement --- instability of the driven branch past a
critical drive \cite{glansdorff1971,prigogine1978} --- which the remainder
of this paper operationalizes for informational driving: the drive is the
ambiguity gradient, the ordered state is the certified macrostate branch,
and the critical drive will be located at unity of a dimensionless ratio
of induced flux to integrative capacity.

\subsection{Information as a Thermodynamic Load}
\label{subsec:info_load}

Information processing is physically instantiated: storage, transformation,
and erasure are state transitions of material degrees of freedom that incur
entropic cost whenever they are logically irreversible
\cite{landauer1961,bennett1982}, independent of the semantic content of the
information involved \cite{shannon1948,cover2006}.

We model this burden as an \emph{informational load}: a rate at which uncertainty
is induced into a system by measurements, signals, coordination demands, or
internal update pressures that must be integrated to maintain coherent
operation. Following Shannon’s formulation of information as uncertainty
reduction \cite{shannon1948}, let $R_{\mathrm{self}}(t)$ denote the induced
informational flux, measured as the rate at which new informational microstates
must be processed, interpreted, or acted upon. Sustained informational flux
implies sustained entropy production, because acquisition, buffering,
coarse-graining, and eventual erasure of information are physically irreversible
operations that dissipate free energy into the environment
\cite{landauer1961,bennett1982}.

Artificial intelligence is not abstract ``software'' atop a substrate:
any physically realizable computation is state transitions of matter and
energy, constrained by finite capacity, dissipation, and
irreversibility \cite{lloyd2000,lloyd2002}; erasure, commitment,
buffering, and coarse-graining are physical events, and the quantitative
form of their entropic floor is developed where it is used
(Sec.~\ref{subsec:sr_entropy}).

Any physical system tasked with integrating informational flux possesses a
finite \emph{integrative capacity} $C_{\mathrm{self}}(t)$, determined by its
substrate, architecture, temporal constraints, and validation mechanisms.

\paragraph{Derivation of the pair $(R_{\mathrm{self}},C_{\mathrm{self}})$
from entropies.}
Neither rate is an independent postulate; both are entropies over
characteristic times --- one environmental, one internal. On the drive
side, the environment presents candidate distinctions at mean spacing
$\tau_{\mathrm{upd}}$, and what a decision must resolve is not the full
microstate entropy of $X$ but the entropy $H_{\mathrm{rel}}$ of its
\emph{action-relevant partition}: the Gibbs entropy of the environmental
ensemble coarse-grained to the distinctions on which commitments
condition. $H_{\mathrm{rel}}$ is thereby indexed to the task: the load
a decoder carries is defined relative to the distinctions its
commitments must resolve, exactly as the hypothesis set defines a
detection problem \cite{vantrees1968detection} --- which is what the
admissible-proxy freedom of Sec.~\ref{sec:rsc_constraint} formalizes.
The induced flux is then the source entropy rate
\cite{shannon1948},
$R_{\mathrm{self}}=\dot H_{\mathrm{env}}
=H_{\mathrm{rel}}/\tau_{\mathrm{upd}}$ --- uncertainty presented per
unit time. On the capacity side, let the representational repertoire be
an internal ensemble of macrostates $z\in Z$ with Gibbs--Shannon entropy
$H_{\mathrm{self}}=-\sum_{z}p(z)\ln p(z)\le\ln|Z|$
\cite{vanrooyen2026} --- the log-measure of distinctions the decoder can
physically hold. A certification cycle registers environmental
distinctions in this repertoire, and Shannon's channel-capacity bound
caps what one cycle can certify: no channel conveys more than $\ln|Z|$
nats per use \cite{shannon1948,cover2006}. With cycles completing no
faster than the validation latency $\tau_{\mathrm{cert}}$, the maximum
certifiable entropy-reduction rate --- the $\dot H_{\mathrm{cert}}$ of
the ambiguity-gradient definition --- is
\begin{equation}
C_{\mathrm{self}}
=\frac{H_{\mathrm{self}}^{\max}}{\tau_{\mathrm{cert}}}
=\frac{\ln|Z|}{\tau_{\mathrm{cert}}}:
\label{eq:cself_derivation}
\end{equation}
capacity is repertoire entropy per validation time, the Shannon capacity
of the decoder viewed as a channel from environmental microstates to
certified macrostates. This is capacity in the elementary
alphabet-bound sense --- trivially achieved by noiseless registration,
only lowered by internal noise, so the impossibility results below are
conservative: saturation against the ceiling implies saturation against
any real certification channel. The genuine coding theorem for
certification --- achievability and converse under a noisy internal
channel --- is posed as the open question of the Conclusion
(Sec.~\ref{sec:conclusion}). The Gibbs
side anchors both physically: each distinguishable macrostate is a
distinguishable substrate state, so $H_{\mathrm{self}}$ is bounded by
the physical entropy available to representation and
$\tau_{\mathrm{cert}}$ by the substrate dynamics
(Appendix~\ref{app:bounds} gives the extreme ceilings). The capacity
ratio follows in dimensionless form,
\[
\mathrm{CR}=\frac{R_{\mathrm{self}}}{C_{\mathrm{self}}}
=\zeta\,\frac{H_{\mathrm{rel}}}{\ln|Z|},
\qquad
\zeta=\frac{\tau_{\mathrm{cert}}}{\tau_{\mathrm{upd}}},
\]
the invertibility timescale ratio of
Sec.~\ref{subsec:local_invertible} weighted by the entropy mismatch
between what a decision must distinguish and what the repertoire can
hold. The minimal model's proxies
$R_{\mathrm{self}}\sim\tau_{\mathrm{upd}}^{-1}$,
$C_{\mathrm{self}}\sim\tau_{\mathrm{cert}}^{-1}$
(Sec.~\ref{subsec:minimal_model}) are the matched unit-entropy case
$H_{\mathrm{rel}}=\ln|Z|$ (one binary certification per candidate), for
which $\mathrm{CR}=\zeta$ exactly --- the coincidence, and its limits,
noted in Lemma~\ref{lem:sufficient_local_invertibility}.

To formalize this balance, we define the \emph{feasibility
margin}
\begin{equation}
\mathcal{M}(t) = C_{\mathrm{self}}(t) - R_{\mathrm{self}}(t),
\label{eq:feasibility_margin}
\end{equation}
which measures the instantaneous slack between induced informational load and
integrative capacity. When $\mathcal{M}(t) \ge 0$, uncertainty can in principle be
managed through delayed commitment, validation, or revision. When
$\mathcal{M}(t)$ is persistently driven toward zero or negative values, linear
adaptation fails and new dissipative regimes become admissible, mirroring the
behavior of nonequilibrium physical systems driven beyond their linear response
range \cite{prigogine1984,deGrootMazur1984}.
Table~\ref{tab:glossary} collects the notation and vocabulary used
throughout.

\begin{table*}[t]
\caption{\label{tab:glossary}Notation and vocabulary. Terms are defined in
full at the cited first use; the admissible-proxy freedom of
Sec.~\ref{sec:rsc_constraint} applies to all rate-valued quantities.}
\begin{ruledtabular}
\begin{tabular*}{\textwidth}{@{\extracolsep{\fill}}lll}
Term (symbol) & Meaning & Introduced \\
\colrule
ambiguity gradient
& \parbox[t]{8.6cm}{\raggedright sustained mismatch $\dot H_{\mathrm{env}}>\dot H_{\mathrm{cert}}$; the imposed nonequilibrium drive}
& Sec.~\ref{sec:physical_framework} \\[3pt]
decoder
& \parbox[t]{8.6cm}{\raggedright any finite-capacity process integrating uncertain observations into irreversible commitments; distinct from the architectural sense (decoder-only transformers)}
& Sec.~\ref{sec:intro} \\[3pt]
$R_{\mathrm{self}}$, $C_{\mathrm{self}}$
& \parbox[t]{8.6cm}{\raggedright induced informational flux $H_{\mathrm{rel}}/\tau_{\mathrm{upd}}$; integrative (certification) capacity $\ln|Z|/\tau_{\mathrm{cert}}$}
& Sec.~\ref{subsec:info_load} \\[3pt]
$H_{\mathrm{self}}$
& \parbox[t]{8.6cm}{\raggedright Gibbs--Shannon entropy of the internal macrostate repertoire; $C_{\mathrm{self}}=H_{\mathrm{self}}^{\max}/\tau_{\mathrm{cert}}$}
& Sec.~\ref{subsec:info_load} \\[3pt]
$\mathcal{M}=C_{\mathrm{self}}-R_{\mathrm{self}}$
& \parbox[t]{8.6cm}{\raggedright feasibility margin; $\Gamma:\mathcal{M}=0$ the feasibility boundary}
& Secs.~\ref{subsec:info_load},~\ref{sec:rsc_boundary} \\[3pt]
$\mathrm{CR}=R_{\mathrm{self}}/C_{\mathrm{self}}$
& \parbox[t]{8.6cm}{\raggedright capacity ratio: the dimensionless drive; open-loop critical value $1$}
& Sec.~\ref{subsec:info_load} \\[3pt]
$SR=\dot N_{\mathrm{uncert}}/\dot N_{\mathrm{cert}}$
& \parbox[t]{8.6cm}{\raggedright stability ratio: order parameter of the feasibility transition; $SR_{\mathrm{EP}}$ its entropy-production twin}
& Sec.~\ref{subsec:sr_entropy} \\[3pt]
synthetic / anchored
& \parbox[t]{8.6cm}{\raggedright entropy produced by uncertified vs.\ certified commitment}
& Sec.~\ref{subsec:sr_entropy} \\[3pt]
$\alpha$ (ungatedness)
& \parbox[t]{8.6cm}{\raggedright mean downstream consumptions of an unverified output}
& Sec.~\ref{subsec:closure} \\[3pt]
$\ell_0$ (grounding), $\theta$ (deadline depth)
& \parbox[t]{8.6cm}{\raggedright exogenous load fraction; decision horizon in service units ($\mu\Delta t$)}
& Sec.~\ref{subsec:closure} \\[3pt]
lucid / collapsed
& \parbox[t]{8.6cm}{\raggedright bounded-backlog stationary branch ($P_{\mathrm{u}}<1$) / self-sustaining fully synthetic branch ($P_{\mathrm{u}}=1$)}
& Sec.~\ref{subsec:closure} \\[3pt]
$q$ (certification gate)
& \parbox[t]{8.6cm}{\raggedright fraction of feedback intercepted; $\alpha_{\mathrm{eff}}=(1-q)\alpha$}
& Secs.~\ref{subsec:local_invertible},~\ref{subsec:thermo_signature} \\[3pt]
spectral overload
& \parbox[t]{8.6cm}{\raggedright the regime $\mathcal{M}(t)<0$: microstates generated and discarded faster than macrostates can be certified}
& Sec.~\ref{sec:intro} \\[3pt]
hallucination regime
& \parbox[t]{8.6cm}{\raggedright sustained $SR\ge SR_c$ (Definition~\ref{def:hallucination_rsc}); a regime label, not a single generative error}
& Sec.~\ref{subsec:ungated_diss_and_synth} \\
\end{tabular*}
\end{ruledtabular}
\end{table*}

\section{Feasibility and Recoverability: A Substrate-Independent Constraint}
\label{sec:rsc_constraint}

The Introduction fixed the question --- classify the operating loop by
exhibiting a critical drive and an instability --- and
Sec.~\ref{sec:physical_framework} fixed the drive. This section chooses
the \emph{object} that carries the analysis, and the choice is dictated
by what must be exhibited: the instability of interest lives where
inference couples to \emph{irreversible action}, so the minimal adequate
object is Shannon's receiver placed under a deadline. We accordingly
model a \emph{decoder} as any physical or informational \emph{process}
that integrates uncertain observations into internal state updates and
downstream commitments, some of which are
irreversible \cite{shannon1948,cover2006}. The name imports its
discipline deliberately: in reliable communication the decoder is the
object for which rate--capacity thresholds are theorems, and this
section states the analog for recoverable action --- feasibility playing
the rate--capacity clause, local invertibility the many-to-one clause
--- with both clauses defined for any substrate, which is what keeps the
constraint substrate-independent. All such systems operate under finite
rate and capacity constraints imposed by substrate, architecture, and
the latency of validation prior to action.

\begin{remark}[Scope: the single-decoder case]
\label{rem:single_decoder}
We treat throughout a \emph{single} decoder that coarse-grains its observed
microstates into one macrostate and takes a hard decision --- the canonical
single-stream case, in the sense that the single-link Gaussian (and, for the
arrival statistics, Poisson) channel is the canonical starting point in
information theory. The realistic generalization is a \emph{joint} decoder
that, rather than committing per stream, estimates the several latent states
--- some estimable within a coherence bandwidth, some independent (diversity),
some mutually correlated --- and resolves the hard decision by joint,
mutual-information detection. Cross-stream agreement then supplies \emph{internal}
certification that a single stream cannot (the error-correcting role noted
in Sec.~\ref{subsec:local_invertible}). This work is deliberately the
single-stream baseline those joint extensions build on
(Sec.~\ref{sec:conclusion}).
\end{remark}

Let $C_{\mathrm{self}}(t)$ denote the decoder’s \emph{effective integrative capacity}
over the relevant decision horizon [derived from the internal entropy in
Sec.~\ref{subsec:info_load}, Eq.~\eqref{eq:cself_derivation}]. Capacity here is not limited to raw computational
throughput; it includes the ability to validate interpretations, certify internal
macrostates prior to action, buffer uncertainty, and preserve coherence under irreversible
updates.

When the feasibility margin [Eq.~\eqref{eq:feasibility_margin}] satisfies
$\mathcal{M}(t)\ge 0$, the decoder can in principle certify internal
macrostates and defer or revise irreversible commitments; as
$\mathcal{M}(t)\to 0$, validation latency and buffering are exhausted; once
$\mathcal{M}(t)<0$ (spectral overload), irreversible actions must occur prior
to certification and recoverability cannot be guaranteed.

\subsection{Minimal model: delayed certification in a finite-state decoder}
\label{subsec:minimal_model}

We now present a minimal, falsifiable model in which all assumptions are
explicit. Over a finite environmental microstate space $X$, the decoder
observes $y_k=h(x_k)$ at discrete times and maps it to an internal macrostate
$z_k$ via a many-to-one compression $f:Y\to Z$; irreversible commitment
occurs when $z_k$ is passed to an action $a:Z\to A$ that eliminates
counterfactual future trajectories in $X$. This is a \emph{detection}
problem in the classical sense \cite{wald1947,vantrees1968detection}
rather than coded communication: the environment is the source and
$h$ an observation channel the decoder does not control, so no cooperative
encoder can be assumed at the source; recoverability must instead be
secured downstream by certifying $z_k$ before commitment. In coded
communication the grounding to the source is agreed in advance, in the
codebook; here it is earned per commitment, by certification --- which
is why certification capacity, not a code rate, is the binding
resource, and why feasibility and local invertibility are the binding
constraints. Certification attempts to
distinguish which equivalence class in $X$ produced $z_k$ prior to action,
with characteristic latency $\tau_{\mathrm{cert}}$; candidate macrostates
arrive with mean spacing $\tau_{\mathrm{upd}}$. Identifying
$R_{\mathrm{self}}\sim\tau_{\mathrm{upd}}^{-1}$ and
$C_{\mathrm{self}}\sim\tau_{\mathrm{cert}}^{-1}$: when
$\tau_{\mathrm{upd}}\gg\tau_{\mathrm{cert}}$, certification completes before
commitment and the mapping from $X$ to committed actions remains locally
invertible; when $\tau_{\mathrm{upd}}\lesssim\tau_{\mathrm{cert}}$,
commitments occur before certification, distinct microstates inducing the
same $z_k$ are acted upon identically, and each uncertified commitment
induces logical irreversibility with strictly positive entropy production.
The model satisfies Assumptions (A1)--(A3) explicitly and admits direct
falsification: certification completing before commitment yet recoverability
failing --- or uncertified commitments occurring without loss of
counterfactual structure --- invalidates the framework.

\begin{definition}[Admissible informational proxies]
The quantities $R_{\mathrm{self}}(t)$ and $C_{\mathrm{self}}(t)$ are defined only up
to an equivalence class of operational proxies evaluated over a decision horizon
$\Delta t$.

An admissible proxy $R_{\mathrm{self}}^{(i)}(t)$ must be monotonically increasing in
the rate at which candidate irreversible commitments are generated over $\Delta t$.

An admissible proxy $C_{\mathrm{self}}^{(j)}(t)$ must be monotonically increasing in
the maximum rate at which such candidates can be certified or validated over the
same horizon.
\end{definition}

All results that follow depend only on the existence of a sustained interval on
which
\[
R_{\mathrm{self}}^{(i)}(t)\;\ge\; C_{\mathrm{self}}^{(j)}(t)
\]
for all admissible proxy choices --- on ordering and divergence properties,
not on absolute calibration or metric choice. If admissible proxies disagree
on the sign of $\mathcal{M}(t)$, the system is by definition operating
arbitrarily close to the feasibility boundary, and the necessity results
apply in the limit $\mathcal{M}(t)\to 0$.

\begin{structclaim}[Recoverability Breakdown Under Sustained High-Flux]
\label{thm:recoverability_breakdown}
Consider a decoder that couples inference to irreversible downstream commitments
and operates under finite effective integrative capacity
$C_{\mathrm{self}}(t)$. Let $R_{\mathrm{self}}(t)$ denote the induced informational
flux over a decision horizon $\Delta t$, and let
$\mathcal{M}(t)=C_{\mathrm{self}}(t)-R_{\mathrm{self}}(t)$ denote the feasibility
margin. Assume:
\begin{enumerate}
    \item[(A1)] Irreversible commitments eliminate counterfactual future options
    (path dependence).
    \item[(A2)] Certification or validation of internal macrostates requires nonzero
    time and finite capacity.
    \item[(A3)] (Accounting convention, physically realizable.) Each internally
    generated state transition that conditions an irreversible commitment
    \emph{prior to completion of certification} is assigned the Landauer cost
    of the certification bit it abandons --- a strictly positive entropy
    production. Appendix~\ref{app:driven_model} exhibits a realization in
    which this assignment is the physical cost; a decoder that defers it
    (e.g.\ by logging all state) meets it at the coarse-graining that finite
    memory eventually forces.
\end{enumerate}
If the induced informational flux satisfies
\[
R_{\mathrm{self}}(t) \ge C_{\mathrm{self}}(t)
\quad\text{over a sustained interval},
\]
then recoverable operation cannot be preserved. Specifically, either:
\begin{enumerate}
    \item[(i)] irreversible actions must be suppressed (no admissible action), or
    \item[(ii)] irreversible actions occur without local invertibility, inducing
    path-dependent loss of future option space.
\end{enumerate}
\end{structclaim}

Assumption (A3) invokes only the minimal Landauer bound, as an accounting
floor: logical irreversibility of the abandoned certification implies
nonzero entropy production, with no assumption on magnitude, rate, or
scaling law.

\paragraph{Scope.}
This structural claim is not a constructive dynamical theorem. It asserts a
regime-level impossibility --- under sustained overload, no admissible policy
can preserve recoverability while continuing irreversible commitment --- and
rules out that \emph{class} of counterexamples without exhausting all
conceivable architectures: it makes no statement about transient, finely
tuned, or externally reset systems outside the stated assumptions.

Feasibility is necessary but not sufficient for recoverable operation.
We now introduce the count-level observable of the feasibility margin---the
stability ratio $SR(t)$---which diverges at the boundary and, there, signals the
capacity-driven loss of local invertibility.

\begin{figure*}[!htbp]
\centering
\includegraphics[width=\linewidth]{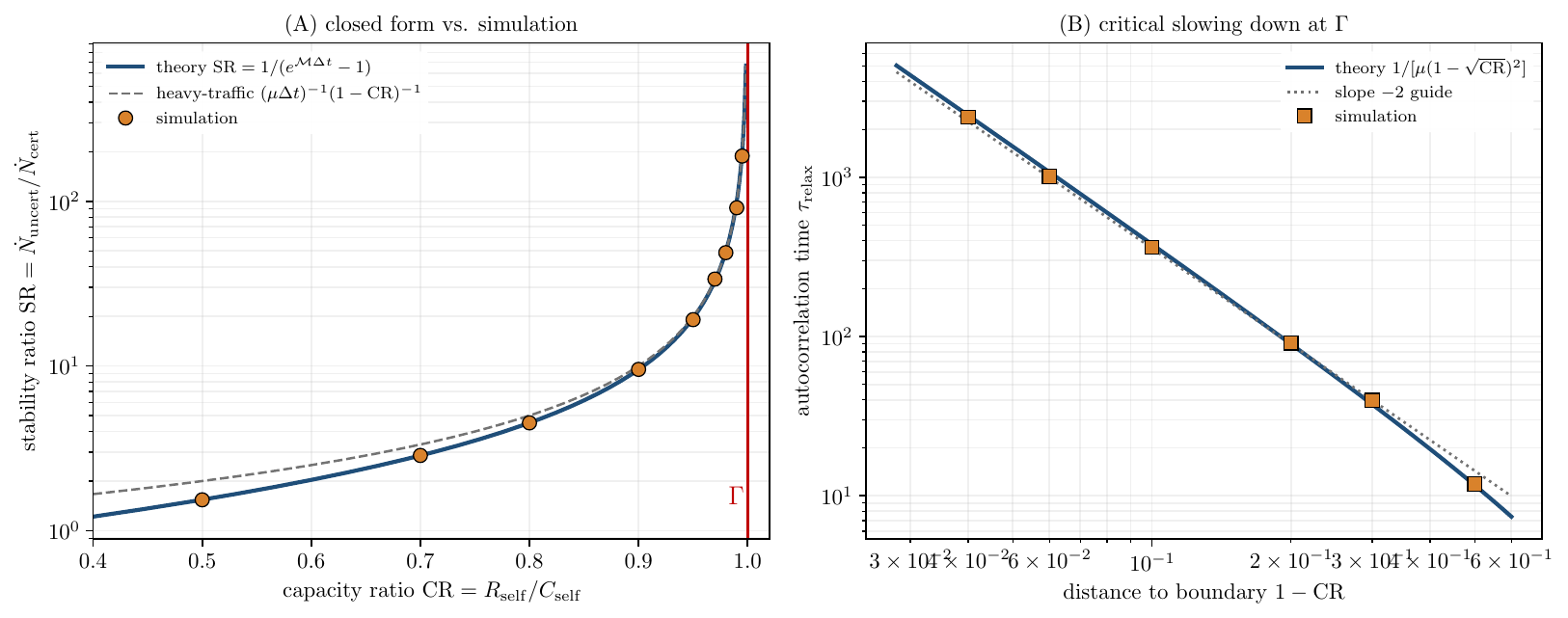}
\captionsetup{width=0.95\linewidth}
\caption{\textbf{Event-driven validation of the closed form and the critical slowing down.}
\textbf{(A)}~The empirical stability ratio (markers) tracks the closed form
$SR=1/(e^{\mathcal{M}\Delta t}-1)$ of Eq.~\eqref{eq:sr_closed_form} (solid), with
heavy-traffic asymptote $(\mu\Delta t)^{-1}(1-\mathrm{CR})^{-1}$ (dashed), across
the approach to the boundary $\Gamma:\mathrm{CR}=1$; the rounding at
$\mathrm{CR}\ge0.99$ is the finite-observation critical window (the
stationary tail requires $T_{\mathrm{obs}}\gg\tau$, panel B).
\textbf{(B)}~The empirical backlog autocorrelation time follows the spectral-gap
prediction $1/[\mu(1-\sqrt{\mathrm{CR}})^2]$, confirming the critical slowing
down $\tau\sim(1-\mathrm{CR})^{-2}$ near $\Gamma$. M/M/1, $\mu=1$,
$\Delta t=1$, $N=3\times10^6$.}
\label{fig:sr_mm1_simulation}
\end{figure*}

\begin{proposition}[Stability-ratio divergence at the feasibility boundary]
\label{prop:sr_divergence}
Model the certification backlog as an M/M/1 birth--death process: candidate
commitments requiring certification arrive first-come-first-served at rate
$\lambda=R_{\mathrm{self}}$ and are certified at rate $\mu=C_{\mathrm{self}}$
(rates in the unit-entropy normalization of Sec.~\ref{subsec:info_load}:
one certification bit per candidate),
so that $\mathrm{CR}\equiv\rho=\lambda/\mu$ and $\mathcal{M}=\mu(1-\rho)$.
A stationary backlog exists iff $\rho<1$ (the feasibility condition itself),
and the FCFS sojourn time is then exactly exponential with rate $\mathcal{M}$
\cite{kleinrock1975}. A commitment fires \emph{uncertified} if its sojourn
exceeds the decision horizon $\Delta t$, so
$\dot N_{\mathrm{uncert}}=\lambda e^{-\mathcal{M}\Delta t}$ and
$\dot N_{\mathrm{cert}}=\lambda(1-e^{-\mathcal{M}\Delta t})$, giving the closed
form
\begin{equation}
SR
\;=\;\frac{\dot N_{\mathrm{uncert}}}{\dot N_{\mathrm{cert}}}
\;=\;\frac{1}{e^{\mathcal{M}\Delta t}-1}
\;\xrightarrow[\;\mathcal{M}\to 0^+\;]{}\;
\frac{1}{\mathcal{M}\,\Delta t}.
\label{eq:sr_closed_form}
\end{equation}
The first equality is the definition of $SR$ as a population ratio
(Sec.~\ref{subsec:sr_entropy}); the class-resolved entropy-production ratio
$SR_{\mathrm{EP}}$ tracks it with a bounded prefactor
(Appendix~\ref{app:closed_loop}).
Equation~\eqref{eq:sr_closed_form} is a Bose--Einstein factor in the single
dimensionless control parameter $\mathcal{M}\Delta t$ (feasibility margin
$\times$ decision horizon). It recovers the heuristic bound
$SR\ge\epsilon/\mathcal{M}$ with the prefactor now identified,
$\epsilon=1/\Delta t$, and diverges as $(1-\mathrm{CR})^{-1}$ at the feasibility
boundary $\Gamma:\mathrm{CR}=1$. The resulting $SR(\mathrm{CR})$ curve
(Fig.~\ref{fig:sr_mm1_simulation}A) is the recoverability analog of a
bit-error-rate curve: $SR$ plays for irreversible action the role $\mathrm{BER}$
plays for reliable communication, and the feasibility boundary
$\mathrm{CR}=1$ is the analog of the Shannon limit --- the rate beyond which
no decoding strategy keeps the error (here, the synthetic-entropy) ratio
bounded.
\end{proposition}

\paragraph{Universality of the exponent.}
The closed form~\eqref{eq:sr_closed_form} is specific to Poisson arrivals and
exponential certification, but the divergence \emph{exponent} is not. In
the heavy-traffic limit, the scaled workload of any GI/GI/1
certification process converges as $\rho\to 1$ to a reflected Brownian motion
with exponential stationary law, so $SR\sim(1-\mathrm{CR})^{-1}$ for every
renewal certification process with finite arrival- and service-time
variances; only the prefactor
carries the arrival/service variability through $(c_a^2+c_s^2)/2$
\cite{kingman1961,whitt2002}. The reversible queue supplies the \emph{rate} of
uncertified commitment; the dissipation that makes $SR$ a thermodynamic ratio is
furnished separately (Sec.~\ref{subsec:sr_entropy}).

\paragraph{Stationarity caveat.}
Equation~\eqref{eq:sr_closed_form} is a stationary statement, while the
queue's relaxation time diverges at the boundary as
$\tau\sim(1-\sqrt{\mathrm{CR}})^{-2}$
(Fig.~\ref{fig:sr_mm1_simulation}B). The divergence is therefore a
quasi-static limit whose approach requires observation windows
$T_{\mathrm{obs}}\gg\tau$, which grow without bound near $\Gamma$: at any
finite $T_{\mathrm{obs}}$ the measured $SR$ rounds off inside a critical
window --- physics, not merely sampling error.

\subsection{Entropy Production and the Stability Ratio}
\label{subsec:sr_entropy}

\paragraph{Physical meaning of internal entropy production.}
Internal entropy production $\dot S_{\mathrm{syn}}(t)$ is defined here through
logical irreversibility rather than microscopic disorder.
Each uncertified internal commitment corresponds to a logically irreversible
state update that erases counterfactual distinctions among environmental
microstates, independent of the physical substrate on which the update is
implemented.

By Landauer’s principle, any such erasure dissipates at least
$k_B T \ln 2$ of heat per lost bit---equivalently, produces at least
$k_B \ln 2$ of entropy---providing a substrate-independent lower bound on
internal entropy production \cite{landauer1961,bennett1982}
(no identification with microscopic thermodynamic entropy beyond this bound is
implied; see Appendix~\ref{app:bounds}).  Accordingly,
\begin{equation}
\dot S_{\mathrm{syn}}(t) \;\ge\; k_B \ln 2 \;\dot{N}_{\mathrm{uncert}}(t),
\label{eq:syn_landauer}
\end{equation}
where $\dot{N}_{\mathrm{uncert}}(t)$ denotes the rate of uncertified internal
commitments over the decision horizon.

\paragraph{Definition of the stability ratio.}
Within the Recoverable Self-Coding (RSC) framework, the \emph{stability
ratio} is defined primarily as a population (count) ratio --- the order
parameter of the feasibility transition:
\begin{equation}
SR(t)\;\triangleq\;
\frac{\dot N_{\mathrm{uncert}}(t)}{\dot N_{\mathrm{cert}}(t)},
\label{eq:sr_thermo}
\end{equation}
the rate of irreversible commitments taken without completed certification
relative to those taken with it. Its thermodynamic content is the one-way
Landauer bound of Eq.~\eqref{eq:syn_landauer}:
every uncertified commitment erases an unresolved alternative at a strictly
positive entropy cost.

Here $\dot S_{\mathrm{syn}}(t)$ and $\dot S_{\mathrm{anc}}(t)$ denote the
\emph{class-resolved} entropy productions of the commitment stream ---
synthetic (produced by uncertified commitment) and anchored (produced by
certified, externally validated commitment). Both are non-negative and their
sum is the stream's total entropy production; they are \emph{not} the
production/exchange pair of the Prigogine balance, whose exchange term is a
signed flux. The entropy-production ratio
$SR_{\mathrm{EP}}\triangleq
\dot S_{\mathrm{syn}}/\dot S_{\mathrm{anc}}$ tracks the order parameter
\eqref{eq:sr_thermo} with a bounded $O(1)$ prefactor in the explicit driven
model of Appendix~\ref{app:closed_loop}, and coincides with it exactly when
the per-event entropy budgets of the two classes are equal; the prefactor
renormalizes the threshold $SR_c$, never the divergence exponent.

An order-of-magnitude remark pre-empts misreading. The floor $k_B\ln2$ per
commitment ($\sim\!3\times10^{-21}\,$J at room temperature) is many orders of
magnitude below the actual dissipation of any physical decoder ---
contemporary accelerators dissipate $\gtrsim\!10^{10}\,k_BT$ per generated
token \cite{horowitz2014}. Nothing in what follows depends on the numerical value of the floor:
its role is structural. Strict positivity and one-way-ness lock the
direction of the synthetic/anchored decomposition; the divergences and
discontinuities are carried by count ratios, with per-event costs entering
only as bounded prefactors. The floor is the substrate-independent component
of the cost; the remainder is substrate overhead.

\paragraph{Where the dissipation comes from.}
One subtlety must be settled, because the M/M/1 backlog of
Proposition~\ref{prop:sr_divergence} is, in its stationary state, a reversible
(detailed-balanced) birth--death chain \cite{kelly1979}: its housekeeping entropy
production is identically zero. The divergence of $SR$ is therefore \emph{not}
the steady-state dissipation of the queue, and no per-event cost diverges
anywhere: the queue sets only the \emph{counting statistics} of the two
commitment classes, while the finite per-event entropy is carried by the
certification event itself --- a \emph{driven} two-level realization of a
single certification produces strictly positive but bounded entropy
production (Appendix~\ref{app:driven_model}). The divergence of
$SR_{\mathrm{EP}}=\dot S_{\mathrm{syn}}/\dot S_{\mathrm{anc}}$ is thus a
\emph{population} effect: finite costs multiplied by a diverging count ratio.
In the explicit construction of Appendix~\ref{app:closed_loop}, where
certification \emph{is} the spin's first passage (so the exponential service
of the queue is derived from the spin dynamics rather than postulated beside
it), both class-resolved productions emerge from one driven population and
$SR_{\mathrm{EP}}$ inherits the exponent of the count ratio with a provably
bounded prefactor. What the thermodynamic layer adds beyond the (reversible)
queueing statistics is therefore specific, not decorative: it fixes the
\emph{sign} --- the one-way-ness of the synthetic/anchored decomposition,
which the detailed-balanced count process alone does not orient --- and it
predicts a \emph{discontinuity} in $SR_{\mathrm{EP}}$ at the fold
(Appendices~\ref{app:closed_loop} and~\ref{app:large_deviations}). The Landauer floor enters only as a bounded
prefactor and is numerically negligible; its role is to lock direction and
supply the entropic order parameter, not to set a scale.

\paragraph{Status of the thermodynamic layer.}
The layer is an oriented accounting, and the paper's claims are
calibrated to that status. Every formal result in this paper --- the closed
forms, the spinodals and cusp, the cascades, the measurements --- is
stated and proven at the count level, and none depends on the numerical
value of the Landauer floor. What the accounting adds is threefold: it
orients the decomposition (the one-way-ness a detailed-balanced count
process cannot supply); it renders the order parameter and its
discontinuity at the fold in entropy units, exactly in the realization
of Appendices~\ref{app:driven_model} and~\ref{app:closed_loop}; and it
carries the layer's own falsifiable content --- the count-level
fluctuation symmetry of the commitment process (the open program of
Sec.~\ref{sec:conclusion}, item~(ii)), measurable on the pipeline of
Sec.~\ref{subsec:experiment}. What it prices is the paper's subject:
option space destroyed before validation, at a strictly positive and
irreducible cost.

Both $\mathcal{M}(t)\ge 0$ and $SR(t)\le SR_c$ are treated as \emph{constraints}
defining admissible action regimes, not optimization targets; the threshold
$SR_c=O(1)$ is a calibrated free parameter of the open-loop framework, not a
universal constant (the closed loop derives it: $SR_{\mathrm{fold}}$,
Sec.~\ref{sec:closed_loop}). $SR(t)$ is a coarse-grained \emph{diagnostic} of operating regime
--- not a microscopic observable, a conserved quantity, or an objective.
Where a normalized margin is needed,
$1-\mathrm{CR}=(C_{\mathrm{self}}-R_{\mathrm{self}})/C_{\mathrm{self}}$; the
rate-valued $\mathcal{M}$ of Eq.~\eqref{eq:feasibility_margin} is retained
throughout. As with $R_{\mathrm{self}}$ and $C_{\mathrm{self}}$, the
class-resolved productions are defined up to equivalence classes of
admissible operational proxies preserving ordering and divergence over the
decision horizon.

\begin{proposition}[Necessity under sustained overload]
\label{prop:sr_necessity}
Under the conditions of Structural Claim~\ref{thm:recoverability_breakdown}, sustained
overload saturates certification capacity, implying that a nonzero fraction of
internally generated state updates must be acted upon without completed local
invertibility certification. Divergence of the stability ratio $SR(t)$ is
therefore not a design choice, control objective, or contingent diagnostic
outcome, but a structural consequence of capacity saturation.

Under sustained overload $R_{\mathrm{self}}\ge C_{\mathrm{self}}$, the
fraction of irreversible commitments taken on uncertified states is bounded
away from zero; by (A3) each contributes strictly positive synthetic
entropy production while anchored production remains capacity-limited, so
$SR(t)$ obeys the closed-form divergence of
Proposition~\ref{prop:sr_divergence} --- and with it $SR_{\mathrm{EP}}$, up
to its bounded prefactor --- independent of any optimization principle,
control policy, or preference over dissipation pathways:
once feasibility margins collapse, loss of recoverability is necessarily
accompanied by dominance of synthetic entropy.
\end{proposition}

\paragraph{Operational interpretation.}
A decoder may continue to generate internally consistent
inferences even as $SR(t)$ rises; what changes is the admissibility of
irreversible action.

Feasibility determines whether a decoder can operate at all; the next question is
whether its operation remains structurally recoverable once irreversible updates
are permitted.

\subsection{Local Invertibility}
\label{subsec:local_invertible}

\begin{definition}[Local invertibility]
\label{def:local_invertibility}
Let $x_t$ denote an environmental microstate, $y_t$ an observation, $z_t$ an internal macrostate, and $a_t$ an irreversible action. A measurement--action regime is locally invertible over a horizon $\Delta t$ if sufficiently small perturbations in $x_t$ induce distinguishable distributions over $z_{t:t+\Delta t}$ prior to irreversible commitment, and if corrective actions remain admissible before $a_t$ eliminates future options.
\end{definition}

Local invertibility is the weakest sufficient condition
ruling out irreversible many-to-one collapse prior to action; it is necessary
only for decoders without external reset or rollback, and alternative
mechanisms (robust redundancy, reversible action sets, error-correcting
commitments) are admissible provided they equivalently preserve
counterfactual structure. The following lemma provides a falsifiable
sufficient condition for its preservation over a finite horizon.

\begin{lemma}[Sufficient condition for local invertibility]
\label{lem:sufficient_local_invertibility}
Let $x_t$ denote the environmental microstate, $y_t$ the observation, $z_t$ the internal macrostate,
and $a_t$ an irreversible commitment taken at time $t_a$.
Let $\tau_{\mathrm{cert}}(t)$ denote the characteristic latency required to certify (validate/ground)
a candidate macrostate $z_t$ against the environment, and let $\tau_{\mathrm{upd}}(t)$ denote the
characteristic time between successive internally generated candidate updates that could trigger
commitment.

Assume that, over a decision horizon $\Delta t$, there exists a measurable certification gate
$q(t)\in[0,1]$ such that irreversible commitment is suppressed whenever certification is not complete.
If there exist constants $\zeta_\star>1$ and $\delta\in(0,1)$ such that, for all $t$ in the operating
interval,
\begin{equation}
\zeta(t)\;\triangleq\;\frac{\tau_{\mathrm{cert}}(t)}{\tau_{\mathrm{upd}}(t)} \;\le\; \zeta_\star
\quad\text{and}\quad
q(t)\;\ge\; 1-\delta,
\label{eq:invertibility_timescale_condition}
\end{equation}
then the measurement--action regime is locally invertible over $\Delta t$ in the operational sense
of Definition~\ref{def:local_invertibility}: by the gate hypothesis, at most a
fraction $\delta$ of irreversible
commitments fire before a certification cycle has completed, so small perturbations in $x_t$
that are resolvable by the certifier remain distinguishable in $z_{t:t+\Delta t}$ prior to
commitment (the lemma's content is the conjunction of two measurable
conditions, not a probabilistic theorem). Under the minimal-model proxies $R_{\mathrm{self}}\sim\tau_{\mathrm{upd}}^{-1}$ and
$C_{\mathrm{self}}\sim\tau_{\mathrm{cert}}^{-1}$, the timescale ratio coincides with the capacity
ratio, $\zeta=\tau_{\mathrm{cert}}/\tau_{\mathrm{upd}}=R_{\mathrm{self}}/C_{\mathrm{self}}=\mathrm{CR}$,
so the gating threshold $\zeta\gtrsim 1$ and the feasibility boundary $\mathrm{CR}\to1$ mark the
same saturation. This coincidence is a feature of the minimal model, not a
general identity --- quantitatively,
$\mathrm{CR}=\zeta\,H_{\mathrm{rel}}/\ln|Z|$
(Sec.~\ref{subsec:info_load}), so the two coincide only under matched
load: the invertibility timescale ratio $\zeta$ and the capacity ratio
$\mathrm{CR}$ are in general distinct control axes---a decoder far below
saturation ($\mathrm{CR}$ small, $SR$ low) can still lose invertibility when the
environmental cause drifts faster than certification tracks it---so that $SR$, the
observable of the capacity axis, leaves the invertibility axis unresolved. The
two-axis instrument separating them (capacity read via $SR$, invertibility via
$\zeta$) extends the open-loop instrument of the companion
paper~\cite{vanrooyen2026proceedings}.

\emph{Falsifiability (measurable test).} Condition~\eqref{eq:invertibility_timescale_condition} is
falsified empirically if the observed fraction of irreversible commitments made without completed
certification exceeds $\delta$ on any sustained interval, or if $\zeta(t)$ is observed to persistently
exceed $\zeta_\star$ while commitments continue to be executed.
\end{lemma}

Local invertibility concerns the \emph{geometry} of the measurement--action
mapping rather than the accuracy of any particular estimate.
When local invertibility holds, inference errors remain \emph{epistemic}:
misinterpretations can be revised through additional evidence, delayed
commitment, or compensatory action. When invertibility fails, however, the mapping
from microstates to macrostates becomes effectively many-to-one. Structural
history is erased prior to validation, and subsequent information cannot restore
option space already eliminated by irreversible commitment.

\paragraph{Feasibility is not accuracy.}
Accuracy concerns correctness of belief; feasibility concerns admissibility
of action --- a decoder may produce accurate posteriors while operating
infeasibly, just as a controller may minimize cost while driving a system
unstable outside its stability margins.

\paragraph{How invertibility enters this paper.}
In the minimal model the two clauses of the constraint are deliberately
fused: completion of certification \emph{is} the
invertibility-preserving event, so invertibility loss enters through the
capacity route alone, and its violation count is precisely what the
stability ratio measures --- Eq.~\eqref{eq:sr_thermo} is, operationally,
the invertibility-violation ratio, and synthetic entropy its per-event
Landauer price. The drift route --- $\zeta$ rising while load stays low
--- is the axis this instrument leaves unresolved (the caveat of
Lemma~\ref{lem:sufficient_local_invertibility}) and the companion's
two-axis instrument separates \cite{vanrooyen2026proceedings}. Condition
(ii) of the constraint stated below
(Sec.~\ref{subsec:statement_recov_const}) therefore does its work here
as the semantic content of the order parameter and of
Observation~\ref{lem:opt_not_recoverable}, not as an independently
varied control; exercising it independently belongs to the joint-decoder
program of the Conclusion (Sec.~\ref{sec:conclusion}).

\subsection{Statement of the Recoverability Constraint}
\label{subsec:statement_recov_const}

\textbf{Recoverability Constraint.}
For any adaptive decoder that couples inference to irreversible downstream
commitments, stable operation requires simultaneous satisfaction of:
(i) a non-negative feasibility margin, $\mathcal{M}(t)\ge 0$, and
(ii) local invertibility of the measurement--action regime.

If either condition is violated, irreversible actions become inadmissible.
Actions taken in such regimes induce path-dependent loss of future option space
that cannot be reversed by subsequent information, capacity increases, or
posterior revision.

\subsection{Irreversibility and Feasibility Boundaries}
\label{sec:rsc_boundary}

The joint behavior of the feasibility margin $\mathcal{M}(t)$ and the stability
ratio $SR(t)$ defines a boundary between recoverable and non-recoverable operating
regimes. We denote the primary feasibility boundary by
\begin{equation}
\Gamma = \{\, \mathcal{M}(t) = 0 \,\},
\end{equation}
which marks the point at which induced informational flux exhausts effective
integrative capacity.

Crossing $\Gamma$ is not a gradual degradation but a qualitative regime
transition. Actions taken beyond this boundary collapse distinctions that would
otherwise support correction or revision, inducing irreversible contraction of
future option space. 

The regime geometry is a plane parameterized by $(\mathcal{M},SR)$:
$\Gamma$ separates feasible from overloaded operation, $SR_c$ marks loss
of admissibility through synthetic-entropy dominance; substrates differ
only in where their operating regimes sit in this plane.

In physical systems, such regime shifts manifest as the emergence of new
dissipative structures. Section~\ref{sec:closed_loop} first closes the loop
that the present analysis leaves open --- the feedback of uncertified
commitment onto the induced load --- and Sec.~\ref{sec:ai_dissipative} then
applies the framework to artificial decoders.

\section{Closed-Loop Decoders: First-Order Recoverability Collapse}
\label{sec:closed_loop}

The preceding analysis treats the induced flux $R_{\mathrm{self}}$ as an
external drive. The framework's own mechanism, however, implies feedback:
commitments taken on uncertified macrostates generate \emph{further} load ---
rework, error-conditioned downstream actions, unvalidated outputs re-entering
the input stream. Systems engineering documents the resulting failure class
at scale as \emph{metastable failures}: self-sustaining congestive collapses
(retry storms, death spirals) that persist after the triggering load is
removed and resolve only under a large corrective action
\cite{bronson2021metastable,huang2022metastable}. That line has begun to
model itself formally --- calibrated queueing and Markov-chain models
that identify metastability and predict recovery for specific systems
\cite{habibi2024msf,alvaro2025formal} --- but what it computes per system
it does not supply as structure: an order parameter, closed-form phase
boundaries, exponents, or a thermodynamic accounting. Congestion transitions are themselves an established
statistical-physics subject: the phase transition in a network-traffic
model entered this journal's literature with Ohira and
Sawatari \cite{ohira1998phase}, congestion onset in communication
networks carries a developed theory \cite{arenas2001communication},
the jamming transition becomes first order under traffic-aware
routing \cite{echenique2005jamming} --- as does packet-level
traffic \cite{hu2007phase} --- and hysteretic transitions in driven
vehicular flow are textbook material \cite{chowdhury2000traffic}.
Those transitions arise from routing, topology, and exclusion; the
mechanism here is deadline-triggered re-consumption of unverified
output, with a closed-form phase diagram in the gating parameters. The
self-sustaining overload also has a classical communications ancestor:
the bistable throughput and congestion collapse of slotted ALOHA under
retransmission feedback \cite{kleinrock1975multiaccess}, the
retransmission counterpart of the deadline mechanism. Analogous error
cascades have recently been reported in multi-agent language-model
pipelines \cite{jamshidi2026hallucination}.
The mathematical literature on retrial queues
\cite{artalejo2008retrial} develops stability conditions
for feedback through an orbit with constant or queue-length-dependent retrial
rates, and deadline-triggered abandonment is likewise classical
\cite{zeltyn2005} --- though there expired jobs leave the queue, while
here they remain enqueued and consume capacity: the no-feedback neighbor
of the present closure. Real pipelines interpolate between the two
ends; the closure models the re-entry end that the metastable-failures
record documents (retry storms persist because expired work re-enters),
and backlog memory --- hence hysteresis --- should weaken toward the
abandonment end, where the workload is bounded: a discriminator
observable in logs. The feedback kernel here differs --- offspring are fired by
\emph{deadline violation}, so the feedback probability is itself
congestion-dependent --- and it is this coupling that produces the fold. This section supplies the
statistical-mechanics account: making the feedback explicit converts the continuous
feasibility transition of Sec.~\ref{sec:rsc_constraint} into a
\emph{first-order} transition with hysteresis, branching cascades, and a
thermodynamic discontinuity. Derivations and simulation methods are
collected in Appendix~\ref{app:closed_loop}.

\subsection{Feedback closure and phase diagram}
\label{subsec:closure}

\begin{figure*}[!htbp]
\centering
\includegraphics[width=0.94\textwidth]{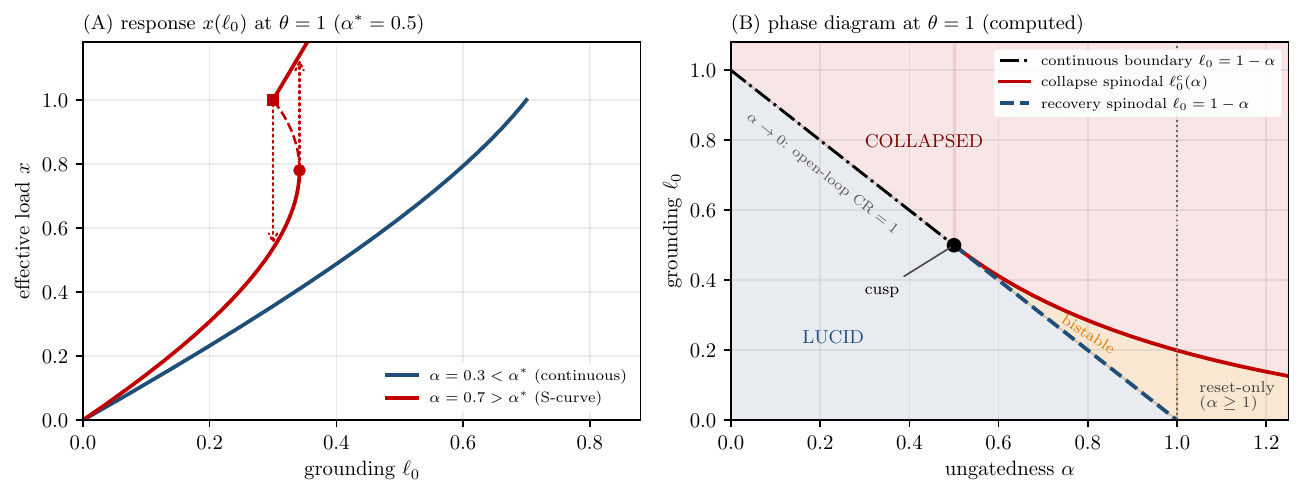}
\caption{\textbf{Closed-loop phase structure}, computed from
Eq.~\eqref{eq:closure} at $\theta=1$
(cusp $(\alpha^{*},\ell_0^{*})=(\tfrac12,\tfrac12)$).
\textbf{(A)}~Response $x(\ell_0)$: below the cusp ($\alpha=0.3$) the
transition is continuous at the renormalized boundary $\ell_0=1-\alpha$;
above it ($\alpha=0.7$) the response is an S-curve --- stable lucid
branch to the fold (circle), unstable middle branch (dashed), and the
collapsed branch $x=\ell_0/(1-\alpha)$, self-sustaining for
$\ell_0\ge1-\alpha$ (square) --- with hysteretic jumps (dotted
arrows). \textbf{(B)}~The $(\alpha,\ell_0)$ plane: collapse spinodal
[Eq.~\eqref{eq:fold_main}], recovery spinodal $\ell_0=1-\alpha$, the
bistable wedge between them, and the continuous boundary below the cusp.
At $\alpha=0$ the open-loop $\mathrm{CR}=1$ boundary of
Sec.~\ref{sec:rsc_constraint} is recovered; for $\alpha\ge1$ the
recovery spinodal reaches $\ell_0=0$ and collapse is irreversible under
load reduction (Prop.~\ref{prop:reset_cure}).}
\label{fig:phase_diagram}
\end{figure*}

Let each uncertified commitment spawn, on average, $\alpha$ new candidate
commitments; $\alpha$ measures \emph{ungatedness} --- the mean number of
downstream consumptions of an unverified output. With exogenous (grounded)
candidates arriving at rate $\lambda_0$ and the stationary uncertified
fraction $P_{\mathrm{u}}(\lambda)=e^{-(\mu-\lambda)\Delta t}$ of
Proposition~\ref{prop:sr_divergence}, the total arrival rate becomes
self-consistent. In dimensionless form
($x=\lambda_{\mathrm{eff}}/\mu$, $\ell_0=\lambda_0/\mu$,
$\theta=\mu\Delta t$),
\begin{equation}
x=\ell_0+\alpha\,x\,e^{-(1-x)\theta} .
\label{eq:closure}
\end{equation}
The closed-loop capacity ratio $x=\mathrm{CR}$ is now a \emph{response}; the
control parameters are the grounding $\ell_0$, the ungatedness $\alpha$, and
the deadline depth $\theta$. The phase structure that follows is a property of
this class of \emph{deadline-feedback} closures --- exogenous grounding plus
Poisson re-consumption of uncertified output fired at the decision horizon.
Its qualitative features (a fold, a cusp, and hysteresis) follow for any
convex, increasing feedback term; the closed-form spinodals below are specific
to the form~\eqref{eq:closure}, which is derived rather than posited (the
$P_{\mathrm{u}}$ factor is the M/M/1 deadline-overflow probability of
Proposition~\ref{prop:sr_divergence}). Equation~\eqref{eq:closure} has an
exact structural precedent: it is the form of the optical-bistability state
equation --- drive plus convex feedback yielding an S-curve with two stable
branches \cite{bonifacio1976cooperative}
[Fig.~\ref{fig:phase_diagram}(A)] --- whose hysteresis loop was the
first-order signature traced experimentally
in \cite{gibbs1976bistability}; the feedback here is supplied by deadline
violation rather than the intracavity field. It is equally the form of
the exothermic-reactor heat balance, whose Arrhenius feedback produces
the ignition/extinction multiplicity of the continuous stirred-tank
reactor \cite{vanheerden1953,uppal1974} --- an engineered driven system
operated against its fold boundaries, with runaway envelopes and quench
procedures, for seventy years; deadline overflow plays the role of
Arrhenius heat release.

Because the feedback term is convex and increasing in $x$,
Eq.~\eqref{eq:closure} has saddle-node structure, and the entire phase
diagram is closed-form (Appendix~\ref{app:closed_loop};
Fig.~\ref{fig:phase_diagram}): a \emph{collapse
spinodal} at
\begin{equation}
\ell_0^{c}=\frac{\theta x_f^{2}}{1+\theta x_f},
\qquad
\alpha(1+\theta x_f)\,e^{-(1-x_f)\theta}=1,
\label{eq:fold_main}
\end{equation}
beyond which no \emph{lucid} state --- the bounded-backlog stationary
state with $P_{\mathrm{u}}<1$ --- exists; a fully synthetic
\emph{collapsed branch} $\lambda_{\mathrm{eff}}=\lambda_0/(1-\alpha)$ with
$P_{\mathrm{u}}=1$, self-sustaining for $\ell_0\ge1-\alpha$ (the
\emph{recovery spinodal}); a bistable wedge between the two; and a
\emph{cusp} at
\begin{equation}
\alpha^{\ast}=\frac{1}{1+\theta},
\qquad
\ell_0^{\ast}=\frac{\theta}{1+\theta},
\label{eq:cusp_main}
\end{equation}
below which the transition is continuous at the \emph{renormalized} boundary
$\ell_0=1-\alpha$. At $\alpha=0$ the open-loop boundary $\mathrm{CR}=1$ of
Sec.~\ref{sec:rsc_constraint} is recovered exactly: the continuous
divergence of Eq.~\eqref{eq:sr_closed_form} is the $\alpha\to0$ edge of a
cusp catastrophe [Fig.~\ref{fig:phase_diagram}(B)]. Note the dependence $\alpha^{\ast}=1/(1+\theta)$: deep
deadlines make even weak feedback first order, because long queues maximize
the congestion coupling.

\begin{proposition}[Collapse pre-empts the divergence]
\label{prop:preempt_main}
On the lucid branch the stability ratio at the moment of collapse is finite:
$SR_{\mathrm{fold}}=1/[\alpha(1+\theta x_f)-1]$. For
$\alpha>\alpha^{\ast}$ the open-loop divergence $SR\to\infty$ is never
traversed; the system jumps discontinuously from a finite
$SR_{\mathrm{fold}}$ (e.g.\ $\approx1.8$ at $\theta=\alpha=1$) to the fully
synthetic state. The closed loop thereby supplies a principled value for the
open-loop admissibility threshold: $SR_{\mathrm{fold}}$ is the derived collapse
ratio that Sec.~\ref{subsec:sr_entropy} could only posit as the calibrated
parameter $SR_c$. \emph{Monitoring corollary:} an $SR$-threshold alarm
calibrated on the open-loop theory waits for a divergence that never
arrives; the operative precursors are the fluctuation signatures discussed
below.
\end{proposition}

\begin{proposition}[Irreversibility at $\alpha\ge1$: the reset cure]
\label{prop:reset_cure}
For $\alpha\ge1$ the collapsed state is self-sustaining at any exogenous
load, and the fluid cascade balance has no finite solution: the candidate
population grows without bound (runaway resource consumption). Load
reduction alone cannot recover the system; the only exits are \emph{gating}
(reducing $\alpha$ itself) or \emph{reset} (externally draining the
backlog). This derives, rather than describes, the engineering observation
that metastable failures resolve only under a strong corrective push
such as load shedding \cite{bronson2021metastable}.
\end{proposition}

\subsection{Cascades: congestion-correlated branching}
\label{subsec:cascades}

The cascade genealogy is a branching process with ratio
$b=\alpha P_{\mathrm{u}}(x^\ast)$, and two exact identities follow from
Eq.~\eqref{eq:closure} on the lucid branch:
$b=1-\ell_0/x^\ast$ and $\langle s\rangle=x^\ast/\ell_0$ --- the mean
cascade size \emph{is} the load-amplification factor. Subcritical branching
theory then gives $P(s)\sim s^{-3/2}e^{-s/s_c}$ with
$s_c\simeq2(x^\ast/\ell_0)^2$: \emph{grounding bounds the cascades}, the
cutoff growing as the grounded fraction of input shrinks. At the fold the
instability factorizes as $b\,(1+\theta x_f)=1$ --- genealogy times
congestion coupling --- so collapse is \emph{not} critical branching
($b_{\mathrm{fold}}<1$); scale-free cascades live only in the
self-referential corner $\alpha\to1^-$, $\ell_0\to0$. The corner is not
self-organized criticality \cite{bak1987}: nothing tunes the system
toward $b=1$; criticality requires externally driving the parameters to
the corner. The $3/2$ class has an empirical instantiation in a
biological decoder: neuronal avalanches in cortical circuits follow the
same subcritical-branching statistics, with measured branching
parameter near unity \cite{beggs2003neuronal} --- in the present
vocabulary, operation near the self-referential corner.

\begin{figure*}[!htbp]
\centering
\begin{minipage}[t]{0.495\textwidth}
\includegraphics[width=\linewidth]{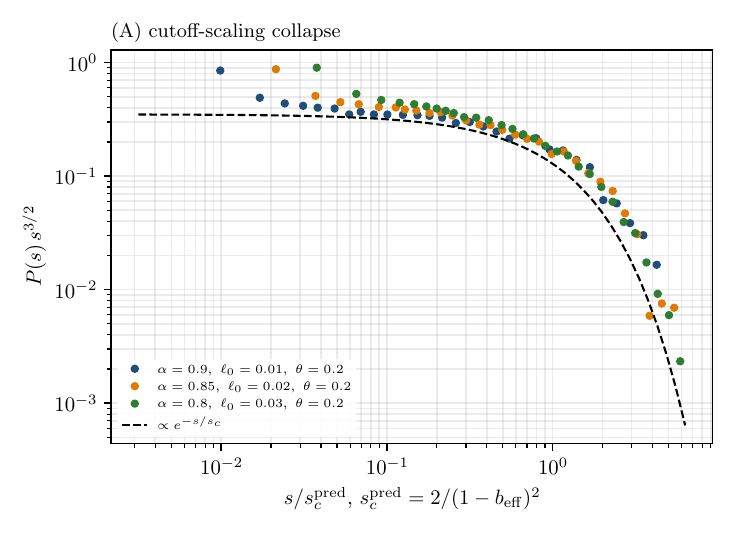}
\end{minipage}\hfill
\begin{minipage}[t]{0.495\textwidth}
\includegraphics[width=\linewidth]{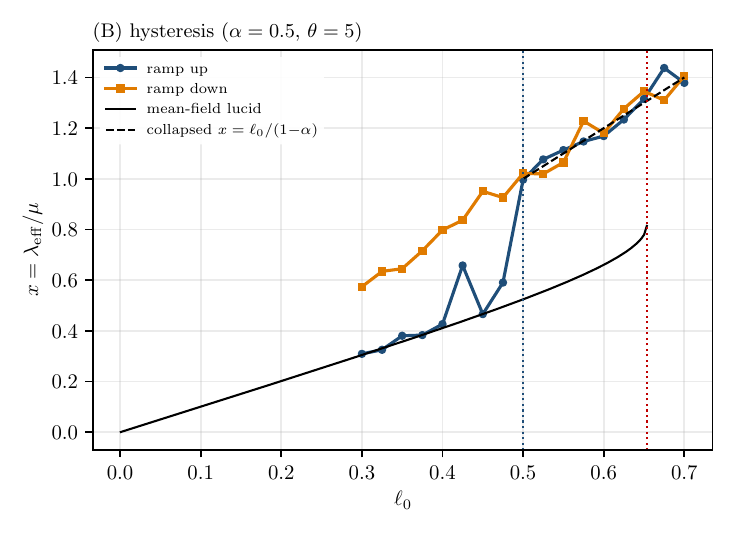}
\end{minipage}
\caption{\textbf{Event-driven simulation of the closed loop.}
\textbf{(A)}~Cascade-size statistics: rescaled distributions
$P(s)\,s^{3/2}$ vs $s/s_c^{\rm pred}$ for three parameter points
($\theta=0.2$; $4$--$12\times10^{4}$ closed cascades each) collapse onto
one master curve with the predicted cutoff
$s_c^{\rm pred}=2/(1-b_{\mathrm{eff}})^{2}$; maximum-likelihood
exponents $\tau=1.502\,[1.494,1.510]$, $1.492\,[1.488,1.500]$,
$1.498\,[1.492,1.506]$ ($1\sigma$).
\textbf{(B)}~Hysteresis ($\theta=5$, $\alpha=0.5$; slow $\ell_0$ ramp,
$1500/\mu$ dwell per step). Blue: up-ramp along the mean-field lucid
branch (solid), with a metastable flicker at $\ell_0=0.425$ and collapse
near $0.50$ --- \emph{below} the mean-field spinodal (red dotted), the
congestion-correlation enhancement. Orange: down-ramp pinned to the
collapsed branch $x=\ell_0/(1-\alpha)$ (dashed) past the recovery
spinodal (blue dotted) by backlog drainage.}
\label{fig:cascade_sim}
\end{figure*}

Event-driven simulations (methods in Appendix~\ref{app:closed_loop})
confirm the exponent and the cutoff law: discrete maximum-likelihood fits
over $0.4$--$1.2\times10^{5}$ closed cascades per parameter point return
$\tau=1.502\,[1.494,1.510]$, $1.492\,[1.488,1.500]$, and
$1.498\,[1.492,1.506]$ ($1\sigma$ profile-likelihood intervals) ---
consistent with the mean-field $3/2$ --- with measured cutoffs tracking
$2/(1-b_{\mathrm{eff}})^{2}$ within a uniform $O(1)$ factor and the
distributions collapsing onto a single master curve
[Fig.~\ref{fig:cascade_sim}(A)].

\subsection{Beyond mean-field: enhancement, hysteresis, and the
sharp-transition limit}
\label{subsec:beyond_mf}

Three simulation results delimit the mean-field theory. First, with zero-lag
feedback, offspring arrive into the very congestion that produced their
parent, so the realized feedback exceeds the mean-field closure; introducing
a spawn latency $\tau_{\mathrm{lag}}$ decorrelates the loop and converges
onto Eq.~\eqref{eq:closure} exactly. \emph{The mean-field phase boundary is
therefore a bound} --- exact in the decorrelated limit, anti-conservative for
tightly coupled loops (measured collapse at $\ell_0\approx0.50$ vs
$\ell_0^{c}=0.653$ at $\theta=5$, $\alpha=0.5$) --- and feedback latency is a
control axis distinct from gating: \emph{delay is a gate}.

Second, the hysteresis loop is directly observed
[Fig.~\ref{fig:cascade_sim}(B)]: the up-ramp rides the lucid branch with a
metastable flicker precursor and jumps near $\ell_0\approx0.50$; the
down-ramp rides the collapsed branch and remains collapsed \emph{below} the
static recovery spinodal, because the accumulated backlog is a memory
variable --- the cost of late intervention grows with time spent collapsed.

\begin{figure}[!htbp]
\centering
\includegraphics[width=\linewidth]{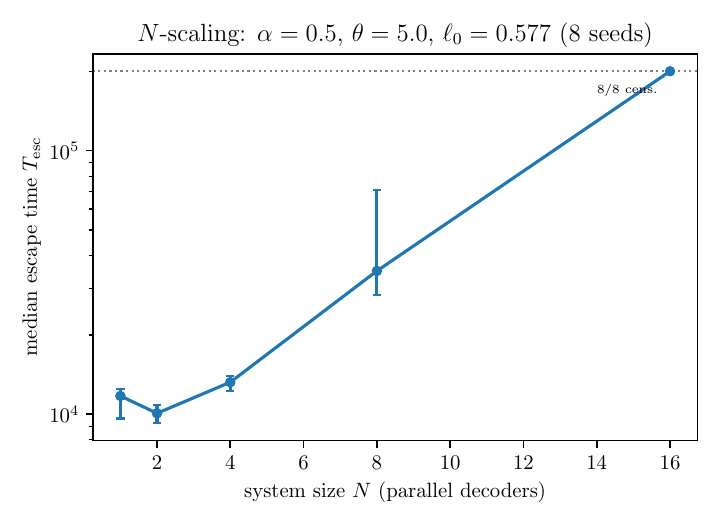}
\caption{\textbf{The fleet limit: lifetime of the lucid state versus
system size.} $N$ parallel decoders share the exogenous stream and the
spawn pool at mid-wedge ($\alpha=0.5$, $\theta=5$, $\ell_0=0.577$;
8 seeds per point, bars spanning the seeded spread). On the logarithmic
axis the median escape time turns upward beyond $N\simeq4$, consistent
with the exponential Freidlin--Wentzell lifetime
$\langle\tau\rangle\asymp e^{NV^{*}}$
(Appendix~\ref{app:large_deviations}); at $N=16$ all eight seeds are
censored at the $2\times10^{5}/\mu$ horizon (dotted). The mean-field
bistability sharpens toward a genuine first-order transition in the
fleet limit.}
\label{fig:nscaling}
\end{figure}

Third, in a single decoder the lucid state inside the wedge is metastable
with short lifetimes ($T_{\rm esc}\sim10^{4}/\mu$ across the wedge;
fluctuation escape, not the deterministic fold, sets the practical
boundary), and deadline depth $\theta$ does \emph{not} sharpen the
transition --- it deepens the wedge but shallows the lucid well. The
sharpening axis is system size: for $N$ parallel decoders sharing the
exogenous stream and the spawn pool, collective fluctuations average as
$N^{-1/2}$, and the measured median lifetime at mid-wedge grows from
$\sim1.2\times10^{4}$ ($N\le4$, 8 seeds each) to $3.5\times10^{4}$ ($N=8$)
to $T_{\rm esc}>2\times10^{5}$ in 8/8 seeds at $N=16$
[Fig.~\ref{fig:nscaling}] --- growth consistent
with the exponential lifetime scaling expected from large-deviation
arguments~\cite{touchette2009large}, under which the mean-field bistability becomes a sharp
first-order transition in the fleet limit. A single tightly coupled
agent is fragile; a fleet is collectively stable, and the corresponding
early-warning program (rising lag-one autocorrelation and variance with
the saddle-node scaling $(\ell_0^{c}-\ell_0)^{-1/2}$ of the local
relaxation rate \cite{scheffer2009early})
is correspondingly reliable only at distance from the fold or in aggregated
fleets, because near the fold the escape can precede converged indicator
estimates.

\subsection{Thermodynamic signature and the price of gating}
\label{subsec:thermo_signature}

The first-order character is visible thermodynamically: sweeping $\ell_0$
across the wedge, the synthetic fraction
$\dot S_{\mathrm{syn}}/(\dot S_{\mathrm{syn}}+\dot S_{\mathrm{anc}})$ jumps
from $\lesssim0.1$ to $\gtrsim0.95$ --- the order-parameter discontinuity
rendered in entropy production --- and each cascade carries a
Landauer-priced burst $\Sigma_{\mathrm{syn}}\ge u\,k_B\ln2$ ($u$ uncertified
members) inheriting the $s^{-3/2}$ tail: synthetic entropy is produced in
scale-free bursts near the self-referential corner. On the collapsed branch
$\dot S_{\mathrm{syn}}\ge k_B\ln2\,\lambda_0/(1-\alpha)$, unbounded for
$\alpha\ge1$. Finally, a certification gate $q\in[0,1]$ acts on the
feedback as $\alpha_{\mathrm{eff}}=(1-q)\alpha$, so exiting the bistable
wedge requires $q>1-1/[\alpha(1+\theta)]$; the gate is itself an information
engine whose operation is bounded by the second law with feedback
\cite{sagawa2008second}, so lucidity carries a calculable minimal
thermodynamic price --- and the certifier, being a finite-capacity decoder,
inherits its own feasibility constraint (cf.\ the no-ideal-decoder remark
closing Sec.~\ref{subsec:design_implications}).

Recursive-training degradation of generative models --- model collapse
under self-consumption \cite{shumailov2024collapse} ---
is the \emph{training-time} sibling of this mechanism: generational
distribution drift, distinct from the inference-time queue saturation
analyzed here in order parameter, timescale, and remedy.

\subsection{Measured collapse in an instrumented computational pipeline}
\label{subsec:experiment}

Following the template of the Introduction, the identification is tested
by measurement on an instrumented physical pipeline: a system whose
timing, service statistics, and scheduling are real, with the feedback
kernel imposed by construction. The rig (code with the seeded protocols
in \cite{cascadecollapse_code}) is a closed-loop computational pipeline:
a single worker performs genuine computation per task (dense linear
solves; empirically calibrated service rate $\hat\mu=1354\,\mathrm{s^{-1}}$ with
coefficient of variation $0.31$ --- markedly non-exponential),
exogenous tasks arrive as a Poisson stream in wall-clock time, deadlines
are wall-clock deadlines ($\Delta t=\theta/\hat\mu$), and a task that
exceeds its deadline fires uncertified at that instant --- spawning
$\mathrm{Poisson}(\alpha)$ genealogy-tagged offspring into the same
queue while continuing to consume real service capacity. Service-time
statistics, dispatcher latency, and operating-system scheduling noise are
all real; the feedback kernel, by contrast, is imposed by construction ---
deadline-triggered firing with $\mathrm{Poisson}(\alpha)$ offspring is the
mechanism of Eq.~\eqref{eq:closure}, not a discovery of the experiment ---
so what the rig tests is the reduction's robustness to everything the
model idealizes away: the service law, real time, and scheduling. The
fold location is not universal (it enters through the service-dependent
overflow probability), so the appropriate mean-field reference below is
computed from the measured service distribution.

\begin{figure*}[t]
\centering
\includegraphics[width=\textwidth]{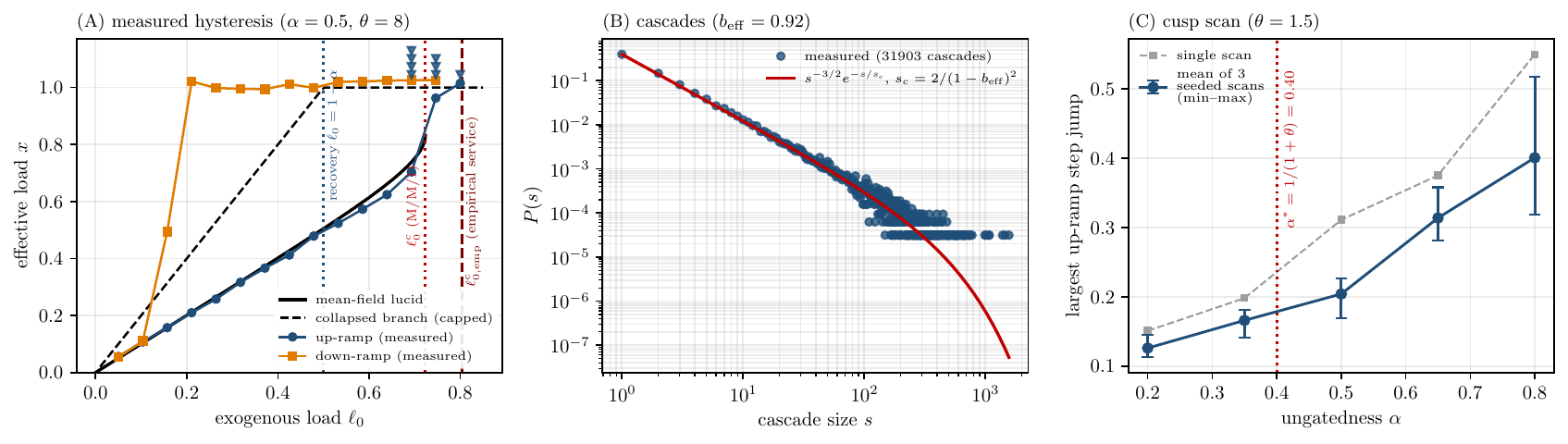}
\caption{\textbf{The real-pipeline experiment.} A single worker performs
genuine computation per task (calibrated $\hat\mu=1354\,\mathrm{s^{-1}}$,
service CV $=0.31$; wall-clock deadlines; real-time deadline feedback).
\textbf{(A)}~Measured hysteresis at $(\alpha,\theta)=(0.5,8)$; the shown
ramp is one of eight seeded realizations, whose first-collapse points
(triangles) scatter up to $14\%$ below the mean-field spinodal computed
from the measured service law ($\ell_{0,\mathrm{emp}}^{c}=0.803$, dark
dashed) and never above it; the exponential-service value $0.724$ is
shown dotted for reference. The down-ramp remains on the collapsed branch far
below the static recovery spinodal $1-\alpha$ while the backlog drains
(the collapsed branch is drawn capped at the service ceiling $x=1$,
where the saturated pipeline pins its measured effective load; the
uncapped fluid branch is $\ell_0/(1-\alpha)$,
Fig.~\ref{fig:cascade_sim}(B)).
\textbf{(B)}~Closed-cascade size distribution at
$(\alpha,\ell_0,\theta)=(0.92,0.15,1.5)$ against
$s^{-3/2}e^{-s/s_c}$ with $s_c$ fixed by the measured mean size.
\textbf{(C)}~Cusp scan at $\theta=1.5$: largest up-ramp step jump in $x$
versus $\alpha$ --- mean over three seeded repetitions (min--max bars)
with a single-scan realization in grey --- growing steeply through
$\alpha^{*}=0.40$.}
\label{fig:experiment}
\end{figure*}

Three measurements (Fig.~\ref{fig:experiment}). \emph{(i) Collapse and
hysteresis.} Ramping the exogenous load $\ell_0$ up and then down at
$(\alpha,\theta)=(0.5,8)$: across eight seeded realizations the up-ramp
rides the mean-field lucid branch and first collapses at $\ell_0=0.693$
(four runs), $0.746$ (three), or $0.800$ (one) --- median $0.720$. The
down-ramp is pinned to the collapsed branch ($x\simeq1$,
$P_{\mathrm{u}}=1$) far below the static recovery spinodal
$\ell_0=1-\alpha=0.5$, recovering only near $\ell_0\approx0.16$ after
the accumulated backlog (peak $\approx1.8\times10^{4}$ tasks) drains ---
the backlog-as-memory delay of Fig.~\ref{fig:cascade_sim}(B), now exhibited
by a real system. Where does this bracket sit relative to theory? The
exponential-service spinodal of Eq.~\eqref{eq:fold_main} is
$\ell_0^{c}=0.724$; but the mean-field reference must use the
\emph{measured} service law. Replacing the M/M/1 overflow probability in
the closure by $P_{\mathrm{u}}(x)$ computed for an M/G/1 queue with
service times bootstrapped from a re-calibration of the same solver on
the same machine ($4000$ samples, $\hat\mu=1425\,\mathrm{s^{-1}}$, CV $=0.297$, against
the run-time $1354\,\mathrm{s^{-1}}$, $0.31$; Lindley recursion --- fed
exponential service, the identical computation returns $0.726$ against
the analytic $0.724$) moves the spinodal to
$\ell_{0,\mathrm{emp}}^{c}=0.803$ [bootstrap 95\% interval
$0.796$--$0.811$ across service-sampling, Lindley-estimator, and
session-recalibration variability; an independent recalibration
returns $0.802$ (protocol in \cite{cascadecollapse_code})]: the
thinner-than-exponential sojourn
tail weakens the deadline feedback at given load. The eight measured
collapse points scatter up to $14\%$ \emph{below} this boundary (median
$10\%$), never above it, the highest-surviving realization ($0.800$)
lying within the boundary's own interval --- the fluctuation-escape
signature of a metastable branch
approaching its fold, and the anti-conservative ordering of
Sec.~\ref{subsec:beyond_mf}, whose zero-lag exponential-service
simulation collapsed $23\%$ below its own boundary at $\theta=5$; the
smaller enhancement here is consistent with the lower service
variability (CV $=0.31$ against $1$), which damps the congestion
excursions that feed offspring back into the queue. The
numerical proximity of the naive M/M/1 value to the measured bracket is
thus partially compensatory --- a thinner service tail raising the true
boundary while feedback correlations and metastable escape advance the
collapse --- and is not claimed as a confirmation of the
exponential-service closed form. \emph{(ii) Cascades.} At
$(\alpha,\ell_0,\theta)=(0.92,0.15,1.5)$, $31{,}903$ closed genealogies
give a size distribution consistent with
$P(s)\sim s^{-3/2}e^{-s/s_c}$ over three decades. The discrete
profile-likelihood estimator used for the simulations returns
$\hat\tau=1.492\,[1.486,1.500]$ ($1\sigma$; $s\ge2$, $19{,}216$
cascades) --- the mean-field $3/2$ --- with fitted cutoff
$\hat s_c=307\,[293,321]$ against the parameter-free prediction
$2\langle s\rangle^{2}=327$ from the measured mean size
$\langle s\rangle=12.796$: agreement within $7\%$. The measured
genealogical ratio $b_{\mathrm{eff}}=1-1/\langle s\rangle=0.922$
marginally exceeds the mean-field ceiling
$\alpha P_{\mathrm{u}}\le\alpha=0.92$ --- the feedback-correlation
enhancement of Sec.~\ref{subsec:beyond_mf}, visible at the genealogy
level. Isolated single-count
events beyond the cutoff are at the sampling floor. \emph{(iii) The cusp.}
Up-ramps across $\alpha$ at $\theta=1.5$, three seeded repetitions: the
largest single-step jump in $x$ grows monotonically through the
mean-field cusp $\alpha^{*}=1/(1+\theta)=0.40$ --- from $0.13$ (spread
$0.11$--$0.14$) at $\alpha=0.2$ to $0.40$ ($0.32$--$0.52$) at
$\alpha=0.8$, the seed spread small against the trend; at this ramp
resolution the crossover from steep
continuous to discontinuous cannot be sharply localized --- a finer
protocol is the immediate refinement.

Scope: this is a single-worker computational pipeline
with an imposed feedback kernel --- the first rung of the measurement
program, the analog of the tabletop sodium cell rather than of the
production device. The completion is the same protocol on
production-scale agentic pipelines, where the re-consumption of
unverified output $\alpha$ is an emergent property of the deployment
rather than an injected parameter, the exogenous rate $\ell_0$, the
certification coverage $q$, and the deadline depth $\theta$ are
measurable configuration, and every observable --- certified and
uncertified commitments, genealogy --- is countable in logs. The
genealogy determines not only the scalar $\alpha$ but the feedback
kernel's \emph{form} --- re-consumption conditional on congestion
state --- directly testing the convex-increasing property on which the
phase structure rests (Sec.~\ref{subsec:closure}).

\section{Illustrative Application: Artificial Decoders}
\label{sec:ai_dissipative}

\paragraph{Scope of application.}
The analysis in this section is illustrative and interpretive; it introduces no
new formal results and does not extend the feasibility or recoverability claims
beyond those established in Sections~\ref{sec:rsc_constraint}
and~\ref{sec:closed_loop}.

The class those sections delimit is deliberately broad. Any
deadline-feedback service system satisfies the closure of
Sec.~\ref{subsec:closure} --- a web tier in a retry storm, a call
center re-queuing its callbacks, a build farm re-running flaky jobs ---
which is why the metastable-failures record reads above as instances
rather than analogies. Artificial decoders are the motivating instance
on two grounds: deployed generative pipelines realize the largest and
least-gated re-consumption of unverified output in operation ---
autoregressive output re-enters context as established input, agentic
systems consume their own artifacts --- and every control parameter of
the phase diagram ($\alpha$, $\ell_0$, $\theta$, $q$) is measurable
configuration in their logs (Sec.~\ref{subsec:experiment}).

Artificial intelligence provides a concrete, contemporary example of a
decoder operating deep within the high-flux regime analyzed above. Throughout
this section, \emph{intelligence} denotes representational and inferential
capacity (mapping observations to macrostates, compressing, and manipulating
them), while \emph{agency} denotes the admissible coupling of those
macrostates to irreversible commitments under the feasibility and
invertibility constraints.

\begin{observation}[Representational optimality does not imply recoverability]
\label{lem:opt_not_recoverable}
Let $X$ denote environment microstate, $Y$ denote observations, and $Z$ denote an internal macrostate produced by a decoder (e.g., $Z=f_\theta(Y)$). Suppose $Z$ is optimized for predictive or inferential performance under a loss $L(\theta)$ (e.g., negative log-likelihood, cross-entropy) so that $\theta^\star=\arg\min_\theta \mathbb{E}[L(\theta)]$. Then, in general, optimality of $Z$ for $L$ does \emph{not} guarantee local invertibility of the measurement--action regime nor recoverability under irreversible commitment.

\emph{Sketch.} Loss-optimal representations are generically \emph{many-to-one}: microstates equivalent for prediction under $L$ map to the same $Z$, which is precisely the failure of local invertibility. Improving accuracy (lower $L$) therefore does not prevent irreversible actions conditioned on $Z$ from eliminating counterfactual structure unless an explicit certification mechanism enforces invertibility prior to commitment.
\end{observation}

\begin{observation}[Scaling laws optimize loss, not recoverability]
\label{cor:scaling_loss_not_rsc}
Empirical scaling laws for large models describe monotonic improvement in predictive
loss as a function of model size, data, and compute \cite{kaplan2020},
but they are silent with respect to recoverability under irreversible action.
Within the Recoverable Self-Coding (RSC) framework, loss minimization improves
inference accuracy (reducing epistemic error) while leaving feasibility margins
and local invertibility constraints unchanged.

Absent explicit certification, higher-throughput generation therefore
drives $SR(t)$ toward criticality rather than restoring recoverability
(Sec.~\ref{sec:rsc_constraint}).

Bayesian optimality characterizes belief updating under uncertainty assuming
reversible inference, whereas Recoverable Self-Coding characterizes action
admissibility under irreversible commitment when the measurement--action loop
itself becomes non-invertible; the latter can fail even when the posterior is
accurate.
\end{observation}

\subsection{Ungated Dissipation and Synthetic Entropy Production}
\label{subsec:ungated_diss_and_synth}

The characteristic failure mode of high-throughput decoding without explicit certification is \emph{ungated dissipation}: internal macrostates are generated and acted upon faster than they can be structurally validated. In RSC terms, this corresponds to elevated stability ratio $SR(t)$, indicating that entropy production is increasingly driven by internal, non-invertible transitions rather than by externally induced uncertainty. The decoder remains active and inference continues, but the measurement--action loop loses local invertibility. Large language models and related architectures are salient instances: high-throughput decoders operating near maximal ambiguity, where coarse-graining is necessarily aggressive.

\begin{definition}[Hallucination as certification-rate violation]
\label{def:hallucination_rsc}
Within the Recoverable Self-Coding (RSC) framework, \emph{hallucination} denotes an
operating \emph{regime} characterized by dominance of uncertified internal
irreversible transitions.

Formally, a decoder is in a hallucination regime over a decision horizon
$\Delta t$ if the stability ratio of Eq.~\eqref{eq:sr_thermo} exceeds an
admissibility threshold, $SR(t)\ge SR_c$,
over a sustained interval of nonzero measure (equivalently, sustained
dominance of $\dot S_{\mathrm{syn}}$ over $\dot S_{\mathrm{anc}}$ up to the
bounded prefactor of Appendix~\ref{app:closed_loop}). Once the closed-loop
parameters $(\alpha,\theta)$ of Sec.~\ref{sec:closed_loop} are measured,
the threshold ceases to be free: $SR_c=SR_{\mathrm{fold}}$ of
Proposition~\ref{prop:preempt_main}.

\end{definition}

The definition is purely structural and thermodynamic --- irreversible
commitments conditioned on macrostates that have not completed external
certification --- and does not presuppose semantic error, model failure, or
inferential inaccuracy. It is a regime label for the \emph{propagation and
commitment} of unverified content in decoding pipelines, agnostic to the
upstream generative mechanism of any individual confabulation; the
single-pass generative failure colloquially called hallucination is the
input to this regime, not the regime itself. Upstream generation has its
own account --- training and evaluation reward guessing over abstention
\cite{kalai2025hallucinate} --- orthogonal to capacity; the two compose:
incentives set why confabulations are generated, the regime condition
sets when they propagate to irreversible commitment.

Crucially, this failure mode is not eliminated by increasing model size, data
volume, or training duration: scaling reduces epistemic error, but absent
certification gates it accelerates the accumulation of non-invertible
transitions, driving $SR(t)$ upward even while $\mathcal{M}(t)$ remains
nominally positive. Ungated dissipation is a regime-level violation of the
Recoverability Constraint, not a deficiency of inference accuracy ---
the expected consequence of operating deep within the high-flux regime
without mechanisms that enforce certification prior to commitment.

\subsection{Structural Instability of Ungated LLM-Centered Action Pipelines}
\label{subsec:llm_instability}

The closed-loop analysis of Section~\ref{sec:closed_loop} yields this as a \emph{consequence of the theory}, not a conjecture: a high-throughput decoder that feeds uncertified output back onto its own load without explicit certification crosses the fold into discontinuous collapse once the effective ungatedness exceeds the cusp value $1/(1+\theta)$, \emph{even when inference accuracy is high} --- accuracy does not enter the cusp condition. The prediction for artificial decoders is correspondingly conditional: to the extent that a deployed large language model (LLM) operates in this regime --- ungated autoregressive generation conditioning irreversible action under high informational flux --- the same result applies.

Accordingly, all instability claims in this section apply to deployment regimes in which LLM outputs are allowed to condition irreversible action directly, and do not apply to language modeling used as a passive, advisory, or fully gated computational component.

LLMs are optimized for high-throughput autoregressive generation under
extreme informational flux \cite{vaswani2017,brown2020}:
aggressive many-to-one compression of contextual microstates into
fixed-width representations and discrete token macrostates, consistent
with information-bottleneck characterizations of deep networks
\cite{tishby2000,shwartz-ziv2017}. This compression is
statistically efficient but not locally invertible unless invertibility
is explicitly imposed architecturally \cite{behrmann2019}:
distinctions among environmental microstates are erased at generation
time, prior to certification --- immaterial for passive inference,
destabilizing when token-level outputs condition irreversible action even
with well-calibrated posteriors. This is why failure modes such as the
hallucination regime (Definition~\ref{def:hallucination_rsc}),
overconfident misgeneralization, and brittle long-horizon behavior
persist in highly capable models: they are regime violations, not
accuracy failures --- inference improves, admissible action does not.

This claim is falsified if an ungated high-throughput decoder is shown to sustain
irreversible action under sustained high induced flux while maintaining bounded
$SR(t)$ and preserved local invertibility.

\subsection{Design Implications: Gating, Certification, and the Role of LLMs}
\label{subsec:design_implications}

The structural instability of ungated LLMs has direct architectural
implications. If high-throughput generative models are to participate in
systems that take irreversible actions, they must be embedded within regimes
that preserve local invertibility: a certification gate $q(t)$ --- external
verification, delayed action, redundancy, cross-model disagreement,
tool-mediated grounding, human-in-the-loop --- constraining when internally
generated macrostates may condition commitment. Stability need not reside
within the LLM itself: models may operate as fast proposal engines upstream
of commitment while certification is handled by slower, externally anchored
mechanisms --- an architectural problem of regime separation rather than a
race toward larger models. Operationally, preservation of invertibility
requires a separation of time scales captured by
$\zeta(t)=\tau_{\mathrm{cert}}(t)/\tau_{\mathrm{upd}}(t)$
(Sec.~\ref{subsec:local_invertible}); recoverable operation requires
suppression of commitment when $\zeta(t)\gg1$, implemented by
$q(t)=q(\zeta(t))\in[0,1]$ with $q'(\zeta)>0$.

Biological and institutional decoders carry endogenous low-bandwidth
realizations of $q(t)$ (fatigue, time pressure, accountability);
artificial systems do not, so validators, delays, redundancy,
tool-grounding, and human-in-the-loop procedures are their structural
substitutes --- and the closed-loop analysis of
Sec.~\ref{sec:closed_loop} quantifies what they must achieve: holding the
effective ungatedness $(1-q)\alpha$ below the cusp value $1/(1+\theta)$.

\paragraph{No ideal decoder.}
No physically realizable decoder sustains irreversible action while
remaining perfectly recoverable under unbounded informational flux:
finite capacity and nonzero validation latency imply that beyond a
regime boundary distinctions are destroyed prior to certification, and
increasing capacity shifts the boundary without eliminating it. In this
light synthetic entropy is counterfactual structure destroyed
prematurely --- lost option space --- which is why the framework
identifies agency with the management of irreversible loss rather than
with capacity alone: higher throughput destroys counterfactual structure
faster unless certification scales commensurately and, with the feedback
of Sec.~\ref{sec:closed_loop}, accelerates entry into the bistable
regime where collapse is discontinuous and recovery hysteretic.

\section{Conclusion: Recoverable Dissipation as the Stability Criterion}
\label{sec:conclusion}

The problem statement asked what kind of physical object an
artificial-intelligence system is, and demanded the answer by
exhibition: a critical drive, an existence boundary, and the transition
between them, measured on a real system. All three have been delivered
for the \emph{operating loop}: the critical drive $\mathrm{CR}=1$ of the
open-loop constraint, advanced under self-referential closure to a
first-order fold --- at mean-field level, sharpening in the fleet
limit --- with closed-form spinodals and cusp; the existence
boundary of the lucid state, whose loss is discontinuous and whose
recovery is hysteretic; and the transition itself, measured on an
instrumented pipeline together with its cascade statistics. The trained
artifact satisfies none of the clauses --- quenched, storable, correctly
managed as a machine. The split answer of the introduction is thus
established rather than asserted: the operating loop is a driven steady
state of the optical-bistability class --- a dissipative structure in
the informational sense --- and its reliability is governed by regime
physics, not component engineering. Beneath the classification sits the
substrate-independent constraint that carries it: recoverable operation
requires a non-negative feasibility margin and local invertibility of
the measurement--action regime; when either is violated, inference may
continue but irreversible action becomes inadmissible, the loss captured
operationally by sustained elevation of $SR(t)$.
Every established row of the audit stated at the outset
(Table~\ref{tab:ds_audit}) is now discharged by the results cited in
its final column.

Increased capacity is stabilizing only conditionally --- it restores
feasibility margins but not invertibility, so ungated throughput
accelerates synthetic entropy production even as accuracy improves ---
and once the feedback of uncertified output onto load is explicit, the
acceleration acquires the sharp phase structure of
Sec.~\ref{sec:closed_loop}. The continuous divergence of the stability
ratio, the centerpiece of the open-loop
analysis~\cite{vanrooyen2026proceedings}, survives as the $\alpha\to0$
boundary of that phase diagram and is dynamically pre-empted by
discontinuous collapse whenever feedback exceeds the cusp value; the
results give quantitative, falsifiable content --- exponents, spinodals,
early-warning scalings --- to a failure class operational practice has
long recognized \cite{bronson2021metastable}.

\paragraph{Practical consequence.}
The stability ratio $SR(t)$ --- with its fluctuation
precursors (rising lag-one autocorrelation and variance near the fold) --- is a
runtime gauge of how close a high-throughput decision system is to collapse;
consistent with Proposition~\ref{prop:preempt_main}, the precursors, not an $SR$
threshold, are the operative alarm. The cusp condition $(1-q)\alpha<1/(1+\theta)$ is a
corresponding design rule, sizing the certification coverage $q$ a pipeline
needs given its re-consumption rate $\alpha$ and deadline $\theta$ --- and
showing that certification may be supplied by a slower external layer rather
than built into the fast generator. Where feedback is already supercritical
($\alpha\ge1$), the reset-cure result selects the remedy: drain the backlog or
cut the feedback, not merely throttle the load.

\paragraph{The classification is operational.}
Table~\ref{tab:machine_vs_ds} collects what the answer to the problem
statement's question changes in practice. The entries are consequences of
results established above, not analogies: postmortems of regime collapse
correctly find no broken component, and reset --- re-nucleation of the
driven state --- is the derived cure rather than a folk remedy
(Prop.~\ref{prop:reset_cure}); rollback plans must target the recovery
spinodal, not the collapse point, and every interval spent collapsed
lowers it further, since the backlog is a memory variable
[Fig.~\ref{fig:cascade_sim}(B) and Fig.~\ref{fig:experiment}]; threshold
alarms on the order parameter wait for a divergence that a first-order
transition never delivers, so the operative telemetry is the fluctuation
precursors (Prop.~\ref{prop:preempt_main}); scaling is not a safety
strategy, because neither accuracy nor capacity enters the cusp condition
$(1-q)\alpha<1/(1+\theta)$ --- the design budget that matters is the
ratio of certification to generation; and evaluation of the artifact
cannot certify the deployment, because the hazard lives in the closure
--- a certified model in an ungated loop is an uncertified system, so the
regulatory analogue is the operating envelope (margin floors, precursor
monitoring, demonstrated reset capability), not product-safety logic.

\begin{table*}[t]
\caption{\label{tab:machine_vs_ds}Practical consequences of the
classification. The left column is component-reliability practice;
chemical and aerospace engineering already run regime doctrines for
specific machines (runaway envelopes, surge margins) --- the right column
is that practice generalized to informational throughput, which computing
infrastructure currently lacks. The entries apply to the
\emph{operating loop} --- the trained artifact is a machine and is
correctly managed as one. Each right-column entry is grounded in a result
of this paper (Secs.~\ref{sec:closed_loop}
and~\ref{subsec:experiment}).}
\begin{ruledtabular}
\begin{tabular*}{\textwidth}{@{\extracolsep{\fill}}lll}
Dimension & Machine & Dissipative structure \\
\colrule
Failure model
& \parbox[t]{5.6cm}{\raggedright a component broke}
& \parbox[t]{7.2cm}{\raggedright the regime left its existence region; nothing is broken} \\[4pt]
Reliability tool
& \parbox[t]{5.6cm}{\raggedright MTBF, redundancy, component QA}
& \parbox[t]{7.2cm}{\raggedright phase diagram, margin monitoring, load shedding} \\[4pt]
Alarm
& \parbox[t]{5.6cm}{\raggedright error-rate thresholds, health checks}
& \parbox[t]{7.2cm}{\raggedright distance to boundary plus fluctuation precursors (Prop.~\ref{prop:preempt_main})} \\[4pt]
Fix
& \parbox[t]{5.6cm}{\raggedright find and repair the fault}
& \parbox[t]{7.2cm}{\raggedright drain the backlog, cut the feedback, reset (Prop.~\ref{prop:reset_cure})} \\[4pt]
Recovery
& \parbox[t]{5.6cm}{\raggedright just below the failure point}
& \parbox[t]{7.2cm}{\raggedright at the recovery spinodal, lowered further by time spent collapsed (hysteresis)} \\[4pt]
Certification
& \parbox[t]{5.6cm}{\raggedright type-test the artifact, once}
& \parbox[t]{7.2cm}{\raggedright conditional on the operating envelope, per regime} \\[4pt]
More capacity
& \parbox[t]{5.6cm}{\raggedright more safety}
& \parbox[t]{7.2cm}{\raggedright faster approach to the fold unless gating scales with generation} \\[4pt]
Maintenance
& \parbox[t]{5.6cm}{\raggedright prevent wear; the system can be paused}
& \parbox[t]{7.2cm}{\raggedright continuous regeneration; the operating state cannot be stored --- checkpoints persist, the running regime does not} \\[4pt]
Regulation
& \parbox[t]{5.6cm}{\raggedright product-safety logic (defect, recall)}
& \parbox[t]{7.2cm}{\raggedright operating-envelope logic (margin floors, precursor monitoring, reset capability)} \\[4pt]
\end{tabular*}
\end{ruledtabular}
\end{table*}

\paragraph{Where the specification lives.}
The classification sharpens under one further question: where does the
information that specifies a structure reside? Three cases arise --- the
specification classes drawn in Fig.~\ref{fig:spec_map}:
\emph{flux-specified} (Class~1), in the present boundary conditions, so
that reimposing the drive re-forms the state --- the convection cell;
\emph{archive-specified} (Class~2), in a separable record, so that the
structure restores bit-identically from its archive --- the checkpoint;
or \emph{trajectory-specified} (Class~3), only in the trajectory that
built it, with no compact record from
which to rebuild. The dichotomy of the problem statement spans the first
two; a deployed decoder mixes all three --- the artifact archived, the
operating regime flux-specified, the uncertified backlog carried only by
the trajectory --- and the closed-loop results locate the failure
physics in the third component: the backlog is the memory variable
behind the hysteresis, past $\alpha=1$ load reduction no longer restores
the lucid state, and the remaining cures are gating and reset
(Prop.~\ref{prop:reset_cure}). Reset falls back on the archive, and
restores exactly what the archive holds --- trajectory-carried state is
destroyed by the reset that cures the regime, not recovered by it.
Certification is the complementary move made in flight: the conversion
of trajectory-carried commitments into archived ones before collapse
forces the fallback. That, and not capacity, is why gating is the
operative stability lever --- what is certified survives the reset that
recoverability may require.

\paragraph{Open program.}
Three theorems would complete the identification of the operating loop as
a dissipative structure in the informational sense posed by the problem
statement: the operating state ordered only under feasible
throughput~(a), its irreversibility accounting thermodynamically
direction-locked~(b), destabilized at critical informational drive by a
genuine nonequilibrium phase transition~(c).
(i)~Promotion of the large-deviation scaffolding of
Appendix~\ref{app:large_deviations} --- tilted generators, the nonconvex
rate function, the spinodals as large-deviation spinodals --- to a theorem
for the self-consistent process: a rigorous LDP for the state-dependent
certification queue, with uniform control at the fold and in the fleet
limit, the condensation transitions of zero-range processes
\cite{evans2005} the natural
comparison class. (ii)~The fluctuation-theorem structure of
Appendix~\ref{app:large_deviations} promoted to a theorem of the
commitment process: the Gallavotti--Cohen symmetry of forward versus
time-reversed commitment paths \cite{lebowitz1999}, proven at the count
level and measurable
on the pipeline of Sec.~\ref{subsec:experiment}.
(iii)~A thermodynamic-uncertainty-type lower bound on the entropy
production required to keep the exceedance probability of the critical
drive below a stated tolerance over a working horizon, stated for the
driven commitment process of Appendix~\ref{app:driven_model} --- the
count-level price of recoverability. The mean-field and simulation results reported here stand on
their own; the identification of the operating loop as an informational
dissipative structure is \emph{completed}, rather than merely
illustrated, by this program --- and by
its empirical twin, the escalation of the Sec.~\ref{subsec:experiment}
protocol from the single-worker pipeline measured here to production-scale
agentic systems --- where the framework's sharpest falsifiable prediction
is that onset of the hallucination regime
(Definition~\ref{def:hallucination_rsc}) tracks \emph{effective load} (context occupancy
weighted by re-consumption density) rather than raw context length, and
where $\alpha$ is directly measurable: a fabricated token at step $k$
re-enters the context as established fact at step $k+1$, the
uncertified-output feedback of Sec.~\ref{sec:closed_loop} realized
literally.

The broader implication for physics: as energy flux through material
media is limited by transport coefficients and instability boundaries,
informational flux through finite decoders is limited by recoverability
--- the feasibility boundary playing for irreversible commitment the role
the Shannon limit plays for reliable decoding
(Prop.~\ref{prop:sr_divergence}), with the added ingredient that
commitment cannot be deferred until certification completes. The
criterion for evaluating high-throughput adaptive systems is therefore
not capacity, performance, or entropy production, but whether
irreversible dissipation remains recoverable: stability is preservation
of option space, not throughput.

Set against the identifications that established the class, the result
inverts the valence of the tradition it borrows from. In the canonical
cases the ordered state appears beyond the critical drive and is the
object of interest --- coherence above the laser threshold, rolls above
the critical Rayleigh number, oscillations past the Hopf point; the
transition \emph{creates} the structure. Here the ordered state of value,
bounded-backlog certified operation, exists \emph{below} the critical
drive, and the transition destroys it. The proof strategy is that of the
laser and optical-bistability programs --- model with drive,
order-parameter reduction, critical condition, class-identifying
signatures, measurement --- but the conclusions land where the
critical-transitions literature operates \cite{scheffer2009early}:
boundaries to be monitored, precursors to be alarmed on, hysteresis to be
planned around. Engineering has operated driven systems against fold
boundaries before --- exothermic reactors with runaway envelopes
\cite{vanheerden1953,uppal1974}, compressors against surge margins ---
but those doctrines never traveled back into the dissipative-structure
literature, and computing infrastructure, which now operates the largest
driven loops in existence, lacks them entirely. It is also where the
physics-of-organisms program opened from within Prigogine's
school \cite{kondepudi2017dissipative} acquires an informational branch.
Table~\ref{tab:machine_vs_ds} exists because here --- unusually for this
literature --- the audience includes the system's governors.

As the first in a series, this work is scoped to the single decoder
(the single-decoder scope remark of Sec.~\ref{sec:rsc_constraint}). The
principal extension is a \emph{joint}
decoder that estimates the several latent states --- estimable, diversity, and
correlated streams alike --- and resolves commitment by joint,
mutual-information detection rather than per stream. Such joint detection
raises the effective integrative capacity and supplies internal certification
through cross-stream agreement, and --- since the cascade then propagates on
the coupling graph rather than as independent branching --- may shift the
avalanche universality class away from the mean-field $3/2$. It also localizes the present account:
an attention-based model is a soft joint decoder that defers the hard
decision, so synthetic-entropy production concentrates at the soft-to-hard
collapse of each autoregressive commitment --- the point at which a
high-capacity joint representation is forced through a hard-decision
bottleneck. Whether this correspondence sharpens into a genuine capacity
theorem for recoverable action --- a mutual-information characterization of
certification capacity, with achievability and converse --- is the question the
joint formulation poses, and one we do not settle here.

\begin{acknowledgments}
The author acknowledges the use of an AI-based assistant (Anthropic's Claude)
in preparing this manuscript --- specifically for prose editing and for
implementing the simulation and figure-generating code. The theoretical
framework, derivations, analysis, and conclusions are the author's own.
The author declares no competing financial interest.
\end{acknowledgments}

\section*{Data Availability}
The code that produces all numerical results and figures of this work
--- the mean-field bifurcation solvers, the event-driven cascade
simulator, the driven two-level simulations, and the
instrumented-pipeline rig with its seeded protocols --- is openly
available in the cascade-collapse repository \cite{cascadecollapse_code}.
The measured pipeline results underlying
Sec.~\ref{subsec:experiment} --- the eight seeded ramp realizations,
the closed-cascade genealogies, the cusp scans, and the
service-calibration samples --- are archived as result files in the
same repository, and every figure regenerates from the seeded commands
documented there.

\appendix

\section{Physical Upper Bounds on Integrative Capacity}
\label{app:bounds}

The results in the main text depend only on \emph{relative} integrative capacity and feasibility constraints, not on proximity to absolute physical limits. For completeness and scale context, we summarize here established upper bounds on information processing that delimit the physically admissible regime of any decoder.

\paragraph{Clarification on entropy.}
Throughout this work, Shannon entropy measures informational uncertainty,
not thermodynamic entropy. Thermodynamic entropy production enters only
through \emph{physically irreversible} operations (erasure,
coarse-graining, irreversible commitment), and no identification between
the two entropies is made beyond the minimal Landauer bound --- entropy
production is an operational consequence of logical irreversibility, not
a state variable or optimization objective.

For scale context: the Margolus--Levitin theorem bounds the rate of
distinguishable state transitions, $\nu_{\max}\le
2E/\pi\hbar$ \cite{margolus1998} --- $\sim10^{51}$ operations per second
for one kilogram of matter \cite{lloyd2000} --- and the Bekenstein bound
limits information density \cite{bekenstein1981}. Biological and silicon
decoders operate many orders of magnitude below these limits; none of
the feasibility or recoverability results requires operation near them.

\section{Falsifiability of the Framework}
\label{app:successive_dissipation}
\label{subsec:falsifiability}
The framework is empirically falsifiable. It would be invalidated by a sustained
regime in which the induced flux satisfies
$R_{\mathrm{self}}\gtrsim C_{\mathrm{self}}$ over extended intervals, irreversible
actions continue, and yet both the feasibility margin $\mathcal{M}(t)$ and the
stability ratio $SR(t)$ remain bounded without explicit gating or capacity
increase --- showing that recoverability can be preserved independently of
feasibility and local invertibility. Equivalently, it fails if measured
irreversible decision breakdowns are fully explained by accuracy degradation
alone, with no independent role for certification latency, saturation, or
path-dependent option loss. Conversely, empirical correlation between feasibility
collapse, rising synthetic entropy, and loss of recoverability supports the
framework without invoking any particular substrate. A test of these
statements must fix its admissible proxy pair and decision horizon
\emph{in advance}: the equivalence-class freedom of the definitions
licenses ordering claims, not post-hoc proxy selection, which could
otherwise accommodate any outcome.

\section{Driven Two-Level Model: The Per-Event Dissipation}
\label{app:driven_model}

Proposition~\ref{prop:sr_divergence} supplies the \emph{rate} of uncertified
commitment from the certification queue, but the queue is reversible in steady
state (Sec.~\ref{subsec:sr_entropy}) and so cannot itself be the source of
dissipation. Here the dissipation is made explicit in a minimal \emph{driven}
statistical-mechanical model, and shown to be genuine but bounded: it grounds
the per-event Landauer cost without reproducing --- or being needed for --- the
divergence.

Render a single certification as an Ising spin $s\in\{\pm1\}$ in a field $h(t)$
the decoder must resolve, $H(t)=-h(t)s$. Glauber dynamics with local detailed
balance,
\begin{equation}
k_{s\to -s}=\frac{1}{2\tau_0}\bigl[1-s\tanh(\beta h)\bigr],
\qquad
\frac{k_{-\to+}}{k_{+\to-}}=e^{2\beta h},
\end{equation}
relax the magnetization toward $m_{\mathrm{eq}}=\tanh(\beta h)$ on a timescale
$\tau_0$ \cite{glauber1963}. Under a time-dependent protocol the occupation lags
equilibrium and the system never reaches a stationary state, so its total
entropy-production rate
\begin{equation}
\dot S_{\mathrm{tot}}
=k_B\,J\,\ln\frac{k_{+\to-}\,p_+}{k_{-\to+}\,p_-}\;\ge\;0,
\qquad
J=k_{+\to-}p_+-k_{-\to+}p_-,
\end{equation}
is strictly positive whenever $p_s\ne p_s^{\mathrm{eq}}$ --- the genuine, driven
dissipation the reversible queue lacks \cite{seifert2012}.
Resolving $\mathrm{sign}\,h$ is one bit; a commitment taken before resolution
erases the unresolved alternative, dissipating at least $k_BT\ln2$ of heat
(Landauer) --- equivalently $k_B\ln2$ of entropy --- the per-event cost underlying
$\dot S_{\mathrm{syn}}\ge k_B\ln2\,\dot N_{\mathrm{uncert}}$.

\begin{figure*}[!htbp]
\centering
\includegraphics[width=\linewidth]{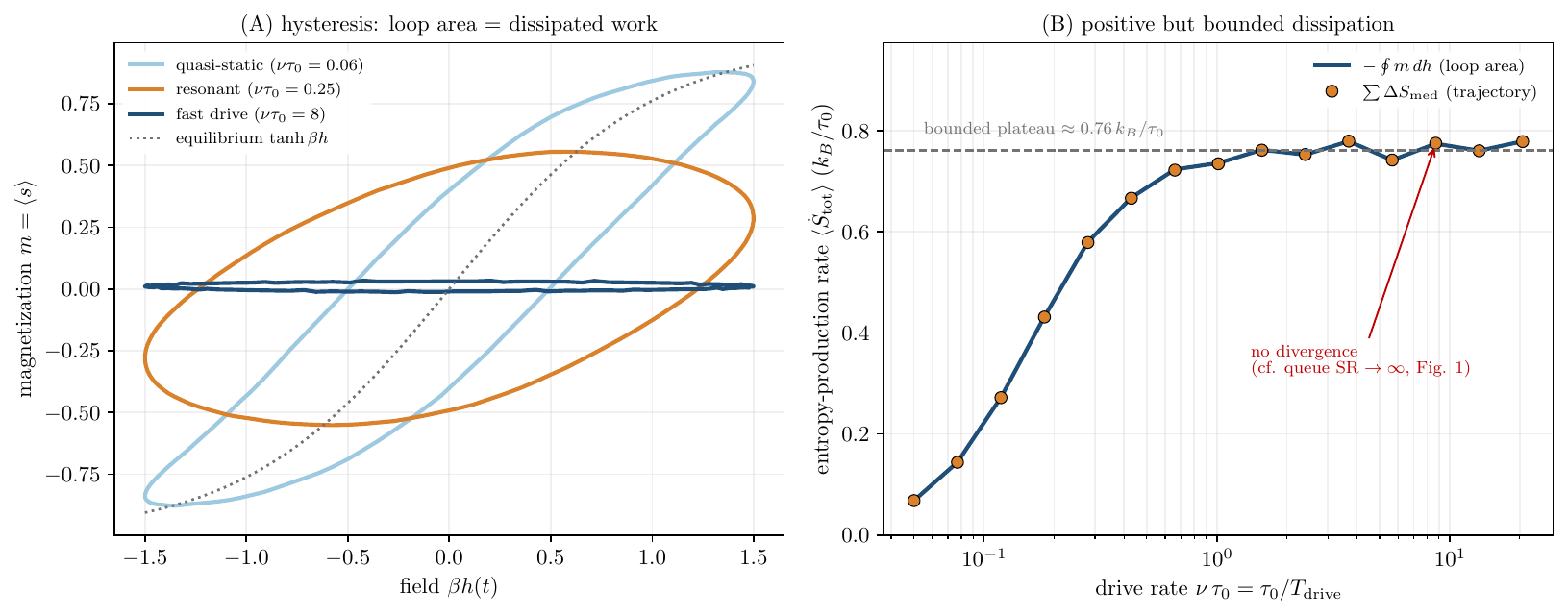}
\caption{\textbf{Driven two-level system: real but bounded dissipation.}
\textbf{(A)}~Magnetization--field loops at three drive rates; loop area equals
the dissipated work per cycle (quasi-static driving hugs the reversible curve;
fast driving collapses the response). \textbf{(B)}~Period-averaged
entropy-production rate versus drive rate, by two independent estimators (loop
area; trajectory medium-entropy sum), rising to a bounded plateau --- positive
everywhere, divergent nowhere, in contrast to the queue
(Fig.~\ref{fig:sr_mm1_simulation}A). $\beta=1$, $\tau_0=1$, $h_0=1.5\,k_BT$.}
\label{fig:ising_dissipation}
\end{figure*}

Crucially, this model is \emph{finite}: its rates and occupations are
bounded, so the single-spin picture supplies the per-event energy scale but
not a divergence --- the latter requires the queue's unbounded backlog
(Proposition~\ref{prop:sr_divergence}).
Figure~\ref{fig:ising_dissipation} confirms both facts by direct simulation
(two independent entropy-production estimators agree to $0.2\%$): dissipation
is strictly positive under driving, vanishes quasi-statically, saturates at
the finite plateau
$\langle\dot S_{\mathrm{tot}}\rangle\to(k_B\beta/\tau_0)\langle h\tanh\beta
h\rangle$, and exceeds the Landauer floor $k_B\ln2$ per resolved bit.

\section{Closed-Loop Bifurcation Analysis and Simulation Methods}
\label{app:closed_loop}

\subsection{Fixed points, fold, and cusp}
With $F(x)\triangleq x-\alpha x e^{-(1-x)\theta}$, steady states of
Eq.~\eqref{eq:closure} solve $F(x)=\ell_0$, and under the natural relaxation
dynamics $\dot x\propto\ell_0-F(x)$ a fixed point is stable iff $F'(x)>0$,
where $F'(x)=1-\alpha e^{-(1-x)\theta}(1+\theta x)$. For
$\alpha(1+\theta)<1$, $F$ is monotone and a unique lucid state exists for
$\ell_0<F(1)=1-\alpha$ --- the continuous boundary. For
$\alpha(1+\theta)>1$, $F$ has an interior maximum at the fold $x_f$ solving
$\alpha(1+\theta x_f)e^{-(1-x_f)\theta}=1$, giving
$\ell_0^{c}=F(x_f)=\theta x_f^{2}/(1+\theta x_f)$
[Eq.~\eqref{eq:fold_main}]; between $\ell_0=1-\alpha$ and $\ell_0^{c}$ the
stable lucid state coexists with an unstable threshold and the collapsed
state. The spinodals merge ($x_f\to1$) at the cusp
[Eq.~\eqref{eq:cusp_main}]. On the collapsed branch the backlog grows
without bound, every commitment times out ($P_{\mathrm{u}}=1$), and the
fluid balance gives $\lambda_{\mathrm{eff}}=\lambda_0/(1-\alpha)$,
self-sustaining iff $\ell_0\ge1-\alpha$; for $\alpha\ge1$ no finite balance
exists (Prop.~\ref{prop:reset_cure}). The use of the stationary sojourn tail
inside the slow load dynamics is an adiabatic approximation, controlled at
any fold with $x_f<1$ since the queue's relaxation time diverges only as
$x\to1$.

\subsection{Branching identities}
Each commitment is uncertified with probability $P_{\mathrm{u}}$ and then
spawns mean $\alpha$ offspring, so the genealogical ratio is
$b=\alpha P_{\mathrm{u}}(x^\ast)$. From Eq.~\eqref{eq:closure},
$\alpha x^\ast P_{\mathrm{u}}=x^\ast-\ell_0$, hence $b=1-\ell_0/x^\ast$ and
$\langle s\rangle=1/(1-b)=x^\ast/\ell_0$ (cascade size = load
amplification; a parameter-free internal check of the simulation, satisfied
to $\sim$1\%). Subcritical Galton--Watson statistics \cite{otter1949} give
$P(s)\sim s^{-3/2}e^{-s/s_c}$, $s_c\simeq2/(1-b)^2$. At the fold,
$F'=0$ reads $b_{\mathrm{fold}}(1+\theta x_f)=1$: the instability is the
product of genealogical branching and the congestion coupling
$\partial P_{\mathrm{u}}/\partial\lambda$, so $b_{\mathrm{fold}}<1$ strictly.

\subsection{Simulation methods}
Event-driven, seeded simulation of the closed loop: exogenous Poisson
arrivals at rate $\lambda_0$ (aggregate $N\lambda_0$ for the fleet); $N$
FCFS servers (default $N=1$) with $\mathrm{Exp}(\mu)$ service; a job whose
sojourn exceeds $\Delta t$ fires uncertified \emph{at its deadline} and
spawns $\mathrm{Poisson}(\alpha)$ offspring (routed uniformly across
servers in the fleet), with an optional exponential spawn latency
$\tau_{\mathrm{lag}}$; jobs remain enqueued until served either way, so
certification capacity is consumed even by already-fired commitments,
matching Eq.~\eqref{eq:closure}. Genealogy tags track cascades to closure.
Validation: at $\alpha=0$ the exact M/M/1 sojourn statistics are reproduced
($P_{\mathrm{u}}=0.606$ vs $e^{-1/2}=0.607$ at $\ell_0=0.5$, $\theta=1$); the
identity $\langle s\rangle=x/\ell_0$ holds to $\sim$1\% at every measured
point; $\tau_{\mathrm{lag}}\gg$ the queue correlation time recovers the
mean-field fixed point ($x=0.2825$ vs $x^\ast=0.2828$ at
$\alpha=0.6,\ \ell_0=0.2$). The mean-field solvers and the event-driven
engine, with the seeded commands that regenerate the figures of this paper,
are openly available~\cite{cascadecollapse_code}.

The per-event entropies use a minimal exact wiring in which certification
\emph{is} a Glauber two-level system's first passage to its field-aligned
state: a two-state chain with an absorbing target has exactly exponential
absorption time, so $\mu=k_+=\tfrac{1}{2\tau_0}[1+\tanh(\beta h)]$ and the
M/M/1 service law is \emph{derived} from the spin dynamics. A certified
event dissipates the finite, load-independent relaxation entropy of the
aligning flip ($\sigma_{\mathrm{anc}}=2\beta h\,k_B$ in this wiring); an
uncertified event erases its unresolved bit at cost
$\sigma_{\mathrm{syn}}\ge k_B\ln2$. Hence
$SR_{\mathrm{EP}}=(\langle\sigma_{\mathrm{syn}}\rangle/
\langle\sigma_{\mathrm{anc}}\rangle)\,SR$ with a prefactor bounded above and
below by load-independent constants: the entropy-production ratio inherits
the order parameter's divergence exponent, and equals it exactly when the
two per-event budgets coincide.

\section{Large-Deviation Structure of the Stability Ratio and the Collapse}
\label{app:large_deviations}

The scaling results of the main text---the closed-form sojourn tail of
Proposition~\ref{prop:sr_divergence}, the first-order fold and cusp of
Sec.~\ref{sec:closed_loop}, and the entropy-production ratio
$SR_{\mathrm{EP}}$ of Appendix~\ref{app:closed_loop}---are not isolated
ans\"atze: each is a special case of, and inherits its rigorous status from,
the large-deviation theory of Markov additive processes
\cite{touchette2009large}. We collect that structure here, both to place the
present results within the established framework of nonequilibrium statistical
mechanics and to fix the route by which the scaling arguments become
derivations.

\subsection{The stability ratio as a Markov additive rate function}
The certification dynamics---the $M/M/1$ backlog of
Proposition~\ref{prop:sr_divergence}, or the queue$\otimes$spin realization of
Appendix~\ref{app:closed_loop}---is a continuous-time Markov process with
generator $\mathcal{G}$, and the stability ratio is built from the
time-averaged certified and uncertified commitment counts over the decision
horizon, i.e.\ from additive functionals
$A_\tau=\tau^{-1}\!\int_0^\tau f(\text{state})\,\mathrm{d}t$. For such
functionals the scaled cumulant generating function
$\lambda(k)=\lim_{\tau\to\infty}\tau^{-1}\ln\langle e^{\tau k A_\tau}\rangle$
is the dominant eigenvalue of the \emph{tilted} generator
$\mathcal{G}_k=\mathcal{G}+kf$, and $A_\tau$ obeys a large-deviation principle
with rate function $I(a)=\sup_k\{ka-\lambda(k)\}$, the Legendre--Fenchel
transform of $\lambda$ \cite{gartner1977,ellis1984,touchette2009large}. Equivalently, the
sojourn-tail exponent that sets $SR$ is the level-1 \emph{contraction} of the
Donsker--Varadhan occupation-measure rate function
$I[\mu]=-\inf_{u>0}\langle\mathcal{G}u/u\rangle_\mu$
\cite{donsker1975,touchette2009large}.
Carrying out that contraction for the birth--death backlog returns precisely
the exponential sojourn tail $\mathbb{P}(T>t)\asymp e^{-\mathcal{M}t}$ used in
Proposition~\ref{prop:sr_divergence}, with decay rate
$\mathcal{M}=\mu(1-\mathrm{CR})$ and the heavy-traffic Kingman exponent as its
general-service form; the closed form $SR=1/(e^{\mathcal{M}\Delta t}-1)$ is
therefore the contracted occupation-measure large deviation, not a postulate.

In this language the open-loop divergence as $\mathrm{CR}\to1$ is the
boundary-singularity (\emph{non-steep}) phenomenon: $\mathcal{M}\to0$ is a
finite-slope boundary point of $\lambda(k)$, which by Legendre duality maps to
an asymptotically linear branch of $I$---an exponential tail whose exponent
vanishes \cite{touchette2009large}. The blow-up of $SR$ is exactly the
vanishing of that exponent.

\subsection{The first-order collapse as a nonconvex rate function}
Closing the loop makes the induced load congestion-dependent
[Eq.~\eqref{eq:closure}], and above the cusp [Eq.~\eqref{eq:cusp_main}] the
stationary backlog density becomes \emph{bimodal}: the lucid fixed point and
the collapsed branch $x=\ell_0/(1-\alpha)$ coexist. The large-deviation image
of this coexistence is a \emph{nonconvex} rate function $I(x)$ with two local
minima, one on each branch \cite{touchette2009large}, and two consequences
follow directly. First, the unstable interior threshold ($F'(x)<0$, the
concave bridge between the minima) is invisible to $\lambda(k)$: the
Legendre--Fenchel transform of $\lambda$ recovers only the \emph{convex
envelope} $I^{\ast\ast}$, so the two spinodals $\ell_0=1-\alpha$ and
$\ell_0^{c}=\theta x_f^2/(1+\theta x_f)$ [Eq.~\eqref{eq:fold_main}], at which
the respective local minima disappear, are the large-deviation spinodals, and
the common-tangent (Maxwell) construction between them is the binodal
\cite{touchette2009large}. Second, the two competing minima are the signature
of broken ergodicity: above the fold the tilted generator $\mathcal{G}_k$ has
two competing dominant eigenvectors, so $\lambda(k)$ acquires a
nondifferentiable kink at the coexistence value \cite{touchette2009large}---the
precise reason the G\"artner--Ellis route fails there and the transition is
first order rather than continuous. The finiteness of $SR_{\mathrm{fold}}$ is
then the statement that the global minimum of $I(x)$ switches
\emph{discontinuously} from the lucid to the collapsed branch while
$\mathcal{M}$ is still strictly positive: collapse pre-empts the divergence
because a first-order jump intervenes before the continuous boundary is
reached.

\subsection{Metastability, hysteresis, and the fleet limit}
The slow load relaxation $\dot x\propto\ell_0-F(x)$, perturbed by the finite
demographic and queueing noise of an $N$-server fleet
(Appendix~\ref{app:closed_loop}), is a noise-perturbed dynamical system of
Freidlin--Wentzell type, with stationary density
$P(x)\asymp e^{-V(x)/\epsilon}$, quasi-potential $V(x)=\inf J[x]$, action
$J[x]=\tfrac12\!\int(\dot x-b(x))^2\,\mathrm{d}t$, and small parameter
$\epsilon\sim 1/N$ set by the fleet size
\cite{freidlin2012,touchette2009large}. In the
bistable wedge each branch is a metastable attractor in its own quasi-potential
well; the collapsed branch persists below the static recovery spinodal for a
mean lifetime governed by the Arrhenius--Kramers escape law
$\langle\tau\rangle\asymp e^{V^\ast/\epsilon}\asymp e^{N V^\ast}$
\cite{freidlin2012,touchette2009large}. The hysteresis of Sec.~\ref{sec:closed_loop} is this
exponentially long metastable lifetime; and because the closed-loop drift is
non-conservative, the up-ramp (fluctuation) and down-ramp (decay) optimal paths
are \emph{not} time-reverses of one another
\cite{freidlin2012,touchette2009large}, which is
the structural origin of the loop's open area. Since the escape barrier scales
with $N$, the metastable lifetime diverges and the bistable wedge sharpens into
a genuine first-order transition as $N\to\infty$: a single tightly coupled
decoder is fragile, a large fleet collectively rigid.

\subsection{The entropy-production ratio as a fluctuation-theorem object}
At the trajectory level the entropy-production ratio
$SR_{\mathrm{EP}}=\dot S_{\mathrm{syn}}/\dot S_{\mathrm{anc}}$ of
Appendix~\ref{app:closed_loop} is the large-deviation entropy production
$W_\tau=\tau^{-1}\ln[\mathbb{P}(\omega)/\mathbb{P}(\omega^{R})]$ of the driven
certification process---the log-ratio of a commitment path $\omega$ to its
time reverse $\omega^{R}$ \cite{touchette2009large}. This is the precise sense
in which $SR$ reads as a ratio of relative entropies between forward and
time-reversed commitment paths. Its generating function obeys the
Gallavotti--Cohen symmetry $\lambda(k)=\lambda(-1-k)$, equivalently the
fluctuation theorem $p(w)/p(-w)\asymp e^{\tau w}$
\cite{lebowitz1999,touchette2009large,seifert2012}: uncertified (irreversible) commitments
are exponentially favored over their time reverses in proportion to the
synthetic entropy they produce---the quantitative form of the one-way-ness that
orients the synthetic/anchored decomposition. Finally, identifying the
synthetic entropy with a rate function and the dimensionless control
$\mathcal{M}\Delta t$ with its conjugate (tilting) field, the discontinuity of
$SR_{\mathrm{EP}}$ at the fold is the nonconcavity of that entropy---and the
kink of the associated free energy (scaled cumulant generating function)---at a
first-order transition: the same convex-analytic signature carried by the
collapse itself \cite{touchette2009large}.

%
%
\bibliography{references}

\end{document}